# PHIBSS: molecular gas content and scaling relations in z~1-3 normal star forming galaxies[1]

L.J.Tacconi[1], R.Neri[2], R.Genzel[1,3,4], F.Combes[5], A.Bolatto[6], M.C.Cooper[7], S.Wuyts[1,] F.Bournaud[8], A.Burkert[9,10], J.Comerford[11], P.Cox[2], M.Davis[4], N.M. Förster Schreiber[1], S.García-Burillo[12], J.Gracia-Carpio[1], D.Lutz[1], T.Naab[13], S.Newman[4], A.Omont[14], A. Saintonge[1], K. Shapiro Griffin[15], A.Shapley[16], A.Sternberg[17] & B.Weiner[18]

[1] *Max-Planck-Institut für extraterrestrische Physik (MPE), Giessenbachstr., 85748 Garching, Germany*
*( linda@mpe.mpg.de, genzel@mpe.mpg.de )*

[2] *IRAM, 300 Rue de la Piscine, 38406 St.Martin d'Heres, Grenoble, France*

[3] *Dept. of Physics, Le Conte Hall, University of California, 94720 Berkeley, USA*

[4] *Dept. of Astronomy, Campbell Hall, University of California, Berkeley, CA 94720, USA*

[5] *Observatoire de Paris, LERMA, CNRS, 61 Av. de l'Observatoire, F-75014 Paris, France*

[6] *Dept. of Astronomy, University of Maryland, College Park, MD 20742-2421, USA*

[7] *Dept. of Physics & Astronomy, Frederick Reines Hall, University of California, Irvine, CA 92697*

[8] *Service d'Astrophysique, DAPNIA, CEA/Saclay, F-91191 Gif-sur-Yvette Cedex, France*

[9] *Universitätssternwarte der Ludwig-Maximiliansuniversität , Scheinerstr. 1, D-81679 München, Germany*

[10] *MPG-Fellow at MPE*

[11] *Department of Astronomy & McDonald Observatory, 1 University Station, C1402 Austin, Texas 78712-0259, USA*

[12] *Observatorio Astronómico Nacional-OAN, Observatorio de Madrid, Alfonso XII, 3, 28014 - Madrid, Spain*

---

[1] *Based on observations with the Plateau de Bure millimetre interferometer, operated by the Institute for Radio Astronomy in the Millimetre Range (IRAM), which is funded by a partnership of INSU/CNRS (France), MPG (Germany) and IGN (Spain).*

[13] *Max-Planck Institut für Astrophysik, Karl Schwarzschildstrasse 1, D-85748 Garching, Germany*

[14] *IAP, CNRS & Université Pierre & Marie Curie, 98 bis Boulevard Arago, 75014 Paris, France*

[15] *Space Sciences Research Group, Northrop Grumman Aerospace Systems, Redondo Beach, CA 90278, USA*

[16] *Department of Physics & Astronomy, University of California, Los Angeles, CA 90095-1547, USA*

[17] *School of Physics and Astronomy, Tel Aviv University, Tel Aviv 69978, Israel*

[18] *Steward Observatory, 933 N. Cherry Ave., University of Arizona, Tucson AZ 85721-0065, USA*

## Abstract

We present PHIBSS, the IRAM **P**lateau de Bure **hi**gh-z **b**lue **s**equence CO 3-2 **s**urvey of the molecular gas properties in normal star forming galaxies (SFGs) near the cosmic star formation peak. PHIBSS provides 52 CO detections in two redshift slices at z~1.2 and 2.2, with $\log(M_*(M_\odot)) \geq 10.4$ and $\log(SFR(M_\odot/yr)) \geq 1.5$. Including a correction for the incomplete coverage of the $M_*$-$SFR$ plane, we infer average gas fractions of ~0.33 at z~1.2 and ~0.47 at z~2.2. Gas fractions drop with stellar mass, in agreement with cosmological simulations including strong star formation feedback. Most of the z~1-3 SFGs are rotationally supported turbulent disks. The sizes of CO and UV/optical emission are comparable. The molecular gas - star formation relation for the z=1-3 SFGs is near-linear, with a ~0.7 Gyrs gas depletion timescale; changes in depletion time are only a secondary effect. Since this timescale is much less than the Hubble time in all SFGs between z~0 and 2, fresh gas must be supplied with a fairly high duty cycle over several billion years. At given z and $M_*$, gas fractions correlate strongly with the specific star formation rate. The variation of specific star formation rate between z~0 and 3 is mainly controlled by the fraction of baryonic mass that resides in cold gas.

*Subject Headings: galaxies: evolution – galaxies: high redshift – galaxies: ISM – stars: formation – ISM: molecules*

# 1. Introduction

The formation and evolution of galaxies involves a complex interplay between the hierarchical merging of virialized dark matter halos, the accretion and cooling of gas onto newly formed galaxies within the halos, the formation of stars in self-gravitating dense gas clouds in these galaxies, and the metal-enriched gas outflows driven by massive stars, supernovae and accreting supermassive black holes in the galaxy nuclei (Rees & Ostriker 1977, White & Rees 1978, White & Frenk 1991, Kauffmann, White & Guiderdoni 1993, Davé, Finlator & Oppenheimer 2011). Recent simulations suggest that most of the build-up of stars in young galaxies, even at the high mass end, may be fed by gradual and semi-continuous gas flows from the cosmic web, or by minor galaxy mergers, plausibly creating gas rich disks (Kereŝ et al. 2005, 2009, Bower et al. 2006, Kitzbichler & White 2007, Ocvirk, Pichon & Teyssier 2008, Dekel et al.2009, Genel et al. 2010, 2012, Governato et al. 2010, Agertz et al. 2011).

Most star forming galaxies between z=0 and 2.5 lie on a near-linear relationship between stellar mass and star formation rate ($SFR \sim (M_*)^p$, p=0.6-0.9 (Schiminovich et al. 2007, Noeske et al. 2007, Elbaz et al. 2007, Daddi et al. 2007, Panella et al. 2009, Peng et al. 2010, Rodighiero et al. 2010, Karim et al. 2011, Salmi et al. 2012, Whitaker et al. 2012). These 'main-sequence' SFGs are typically characterized by disky, exponential rest-UV/rest-optical light distributions ($n_{Sersic} \sim$1-2, Wuyts et al. 2011a). At z>1 main-sequence SFGs double their mass on a typical time scale of ~300-500 Myrs but their growth appears to halt suddenly when they reach the Schechter mass, $M_* \sim 10^{10.8..11} M_\odot$ (Conroy & Wechsler 2009, Peng et al. 2010). This quenching is not yet understood, but evidence is accumulating that star formation, stellar mass or mass surface density thresholds may all play important roles (Kauffmann et al. 2003, 2012, Peng et al. 2010, Bell et al. 2012).

Measurements of mass fractions, spatial distributions and kinematics of cold molecular gas across cosmic time, and the evolution of the relation between gas content and star formation activity, together provide critical observational tests of the 'equilibrium growth' framework described above (e.g. Bouché et al. 2010, Davé, Finlator & Oppenheimer 2012). Star formation in the Milky Way and other local Universe galaxies occurs in massive ($10^4…10^{6.5}$ $M_\odot$), dense (n(H$_2$)~$10^2…10^5$ cm$^{-3}$) and cool (T$_{gas}$~10-50 K), turbulent 'giant molecular clouds' near virial equilibrium on large scales (GMCs: Solomon et al. 1987, Blitz 1993, McKee & Ostriker 2007, Bigiel et al. 2008, Leroy et al. 2008, Bolatto et al. 2008, Krumholz, Leroy & McKee 2011, Kennicutt & Evans 2012). On galactic and sub-galactic scales (0.5 < d <10 kpc) the mean star formation surface density scales linearly or somewhat super-linearly with the molecular gas surface density. This Kennicutt-Schmidt ('KS') – relation, $\Sigma_{star\_form}$~$(\Sigma_{mol\_gas})^N$ , N~1..1.4 (Kennicutt et al. 2007, Bigiel et al. 2008, Leroy et al. 2008, Kennicutt & Evans 2012) has a scatter of ±0.3..0.4 dex. This indicates that the molecular gas depletion time scale, $t_{dep}$= $M_{mol\_gas}$/$SFR$, in normal systems is 1-2 Gyrs, with a mass-dependent variation of about 0.3 dex over $logM_*$=10.4 to 11.4 (Bigiel et al. 2008, 2011, Leroy et al. 2008, Saintonge et al. 2011b).

Improvements in the sensitivity of the IRAM Plateau de Bure millimeter interferometer (PdBI; Guilloteau et al.1992, Cox et al.2011) have recently made it possible to make a census of the molecular gas contents in main-sequence SFGs near the peak of the cosmic galaxy formation activity. Over the last four years we have carried out a CO 3-2 survey of UV/optically selected SFGs in two redshift bins, one at z=1-1.5, and the other at z=2-3. The first results from the survey with 9 <z>=1.2 and 10 <z>=2.2 SFGs and including the first sub-arcsecond CO imaging of a z=1.1 SFG, were reported in Tacconi et al. (2010) and Genzel et al. (2010). Likewise Daddi et al.

(2010a) reported a sample of 6 z~1.5 SFGs. In this paper we report the results of the full CO survey (including the early results), with a total of 38 detected <z>=1.2 SFGs, and 14 detected <z>=2.2 SFGs. For 9 of these SFGs we also have A-and/or B-array sub-arcsecond maps of the spatially resolved CO emission, which are briefly discussed here and in the upcoming papers by Combes et al.(in prep.) and Freundlich et al. (2013).

Throughout the paper, we use the standard WMAP ΛCDM cosmology (Komatsu et al. 2011) and a Chabrier (2003) initial stellar mass function (IMF).

# 2. Observations

## 2.1 PHIBSS Survey Strategy

To explore the cold molecular gas in typical SFGs at z=1-3, we observed the $^{12}$CO 3-2 line emission in samples of SFGs near the star forming main-sequence, selected from large UV/optical/IR look-back imaging surveys. While somewhat biased against dusty and extreme objects, these surveys provide a good census of the 'typical', or 'normal' star forming population at the peak of the cosmic star formation. As described below in more detail, we culled from these surveys SFGs with well determined optical redshifts and with similar ranges in stellar mass and star formation rates in two redshift slices: one at z~1-1.5 ($t_o$=5.4 Gyrs after the Big Bang), the other at z~2-2.5 ($t_0$=3.2 Gyr).

To assure high probability of detection in realistic on-source integration times, we applied a stellar mass cut ($M_*\geq2.5\times10^{10}$ $M_\odot$) and a star formation cut ($SFR\geq30$ $M_\odot$ yr$^-$ [1]). With these cuts, PHIBSS fully covers the massive tail of the stellar mass-star formation plane from the main-sequence line upward, in both redshift slices.

After observing ~15 sources in each redshift bin in the first phase of the survey, we then concentrated on the z=1-1.5 slice for the second stage of the survey, in order to assemble a sufficiently large sample (the initial goal was 35, the final number achieved 38) to study the stellar mass dependence of molecular gas fractions, to determine the slope of the z~1.2 KS-relation, and to increase the number of spatially well resolved cases.

## *2.2 Sample selection*

### 2.2.1 z=1-1.5

The z=1-1.5 main-sequence SFGs selected in this paper were drawn from the All-Wavelength Extended Groth Strip International Survey (abbreviated here as EGS: Davis et al. 2007, Noeske et al. 2007; RA=$14^h17^m$, Dec= $52^030'$). The EGS survey provides deep imaging in all major wave bands from X-ray to radio (including Advanced Camera for Surveys (ACS) HST images) and optical spectroscopy (DEEP2/Keck; Newman et al. 2012) over a large area of sky (0.5 $deg^2$) with the aim of studying the panchromatic properties of galaxies over the last half of the Hubble time. The EGS data provide the properties of a complete set of galaxies from $0.2 \leq z \leq 1.2$, with less complete coverage at higher redshifts, for the stellar mass range >$10^{10}$ $M_\odot$. From the basic EGS data set we selected SFGs with [OII] spectroscopic redshifts in the range z~1.1-1.5, a stellar mass $\geq 2.5 \times 10^{10}$ $M_\odot$ and a star formation rate $\geq 30$ $M_\odot$ $yr^{-1}$. In stage 2, we concentrated on that section of the EGS that is also covered by the recent WFC3 J-H-band coverage in CANDELS (Grogin et al. 2011), as well as by the Hα objective prism observations in 3D-HST (Brammer et al. 2012, van Dokkum et al. 2011), including additional optical spectroscopy from the DEEP3 Galaxy Redshift Survey (Cooper et al. 2011, 2012).  In defining the PHIBSS sample, no morphological selection was applied; targets were selected to sample the complete range of star formation rates within the aforementioned stellar mass limit, ensuring a relatively uniform sampling of the SFR-stellar mass relation at z=1-1.5 (see Fig. 1). As a secondary selection criterion, targets falling within the both the CANDELS and 3D-HST footprints were prioritized, with some consideration given to the redshift of the source so as to enable ground-based follow-up of critical emission lines (such as Hα). The requirement of a high quality optical redshift for the CO follow-up tends to bias the sample to bluer SFGs with relatively low extinction. To test the impact of this bias on the CO properties, we selected in the last phase of the survey two main-sequence galaxies with a significantly larger far-IR excess than for the average of the spec-z population at their mass ($logM_*$~10.7, log($SFR_{IR}$/$SFR_{UV}$)~1.2 instead of 0.75). Their gas fractions do not deviate from the average of the population average.

### 2.2.1 z=2-2.5

The z=2-2.5 SFGs were mainly selected from the near-infrared, long-slit spectroscopy Hα-sub-sample (Erb et al. 2006) of the Steidel et al. (2004) UV-color-magnitude z~2 survey. They were culled from the parent imaging sample with the so-called 'BX/MD' criteria based on UGR colors. Our sub-sample was chosen from the Erb et al. Hα sample to cover the same stellar mass and star formation range as the z~1.2 EGS sample. At the stellar masses and star formation rates considered here, the BX/MD sample provides a reasonably fair census of the entire UV-/optically selected SFG population in this red-shift and mass range (Daddi et al. 2007, Reddy et al. 2005), although the UV-selection is naturally biased against dusty galaxies. The majority of the BX/MD galaxies do not show evidence for undergoing major mergers (Förster Schreiber et al. 2006, 2009, Shapiro et al. 2008). Several of the BX/MD galaxies observed in CO 3-2 were also part of the SINS integral field spectroscopy survey presented in Förster Schreiber et al. (2009) and in the surveys of Law et al. (2009, 2012b), where spatially resolved Hα kinematic data were obtained (for Q1623-BX453, Q1623-BX528, Q1623-BX599, Q1623-BX663, Q2343-BX442, Q2343-BX610, Q2343-BX513, Q2343-BX389, and Q2346-BX482). One galaxy (ZC406690) was taken from the zCOSMOS-SINFONI sample of Mancini et al. (2011, see Genzel et al. 2011) and was selected on the basis of the (s)BzK optical color technique (Daddi et al. 2007).

## *2.3 CO observations and data analysis*

The CO observations were carried out between June 2008 and June 2012 with the 6x15m IRAM Plateau de Bure Millimeter Interferometer (Guilloteau et al. 1992, Cox et al. 2011), as part of two Large Programs. Table 1 summarizes our observations. We observed the $^{12}$CO 3-2 rotational transition (rest frequency 345.998 GHz), which is shifted into the 2mm and 3mm bands for the z~1.2 and z~2.2 sources, respectively. For both bands the observations take advantage of the new generation, dual polarization receivers that deliver receiver temperatures of ~50 K single side band (Cox et al. 2011). For source detections we used the compact 'D' and/or 'C' configurations of the instrument resulting in ~4" and ~2" FWHM beam sizes for observations at 3 and 2 mm wavelength, respectively. We also selected a subsample of 8 z~1.2 SFGs (EGS12007881, 12011767, 13003805, 13004291, 13011166, 1301928, 13026117, 1303518) and one z~2.2 SFG (Q2343 BX610) for high resolution imaging observations in the 'B' and extended 'A' configurations (760 meter maximum baseline), resulting in FWHM beam sizes from 0.3" to 1" in these sources. To maximize the sensitivity and UV coverage of the maps and to ensure that no flux was resolved out at the highest resolution we combined the extended configuration data with the compact configuration observations. The total integration times listed in Table 1 are for all configurations included for a given galaxy.

Weather conditions during the observations varied and data were weighted appropriate to the system temperature (referred to above the atmosphere) at the time of the observations. Depending on the weather conditions and season, system temperatures ranged between 100 and 200 K in both bands. Every 20 minutes we alternated source observations with a bright quasar calibrator within 15 degrees of the source. The absolute flux scale was calibrated on MWC349 ($S_{3mm}$=1.2 Jy). The spectral correlator was configured to cover 1 GHz per polarization. For observations after winter 2010, we also had access to the Widex backend with 4 GHz per polarization. The source integration times were between 4 and 31 hours for the detection sample and 4 to 23 hours for the mapping sub-sample. The data were calibrated using the CLIC package of the IRAM GILDAS software system and further analyzed and mapped in the MAPPING environment of GILDAS. Final maps were cleaned with the CLARK version of CLEAN implemented in GILDAS. The absolute flux scale is better than ±20%. The spectra were analyzed with the CLASS package within GILDAS.

We note that at z=1-1.5 our detection rate actually was >100%. In 2 of the 35 fields targeted toward optically selected SFGs we detected 3 additional CO emitters. These additional sources are SFGs displaced by 50-75 kpc in projection and a few $10^2$ km/s in redshift, either in the same redshift spike, or in a group. The detection rate at z=2-2.5 was somewhat lower (14 out of 20). The lower success rate is partly an expected consequence of the larger luminosity distance at z~2.2 relative to ~1.2. A gas mass of $10^{10}$ M$_\odot$ results in an integrated CO 3-2 flux of 0.042 Jy km/s at the former redshift, and 0.13 Jy km/s at the latter. However, as discussed in Genzel et al. (2012) the faintness or even non-detection of several bright z~2.2 SFGs at first is quite puzzling. In that paper we proposed that the lower than expected CO emission is the result of the more extensive UV photodissociation of CO molecules in lower metallicity environments, leading in effect in an increase of the conversion factor α by factors of >3 and >10 between solar and 0.5 and 0.25 times solar metallicity, respectively (see also Krumholz et al. 2011, Leroy et al. 2011).

## *2.4 Derivation of molecular gas masses*

Observations in giant molecular clouds (GMCs) of the Milky Way have established that the integrated line flux of $^{12}$CO millimeter rotational lines can be used to infer molecular gas masses, although the CO molecule only makes up a small fraction of the entire gas mass, and its lower rotational lines (1-0, 2-1, 3-2) are almost always very optically thick (Dickman, Snell & Schloerb 1986, Solomon et al. 1987). This is because the CO emission in the Milky Way and nearby normal galaxies comes from moderately dense (volume average densities $\langle n(H_2) \rangle$ ~200 cm$^{-3}$, column densities $N(H_2)$~$10^{22}$ cm$^{-2}$), self-gravitating GMCs of kinetic temperature 10-50 K. Dickman et al. (1986) and Solomon et al. (1987) have shown that in this 'virial' regime, or if the emission comes from an ensemble of similar mass, virialized clouds spread in velocity by galactic rotation, the integrated line CO line luminosity $L'_{CO} = \int_{source} \int_{line} T_R \, dv \, dA$ ($T_R$ is the Rayleigh-Jeans source brightness temperature as a function of Doppler velocity $v$) is proportional to the total gas mass in the cloud/galaxy. This total molecular gas mass (including a 36% mass correction for helium) then depends on the observed CO J →J-1 line flux $F_{CO\ J}$, source luminosity distance $D_L$, redshift z and observed line wavelength $\lambda_{obs\ J} =\lambda_{rest\ J}\ (1+z)$ as (Solomon et al. 1997)

$$M_{gas}/M_\odot = 1.75 \times 10^9 \left(\frac{\alpha}{\alpha_G}\right)\left(\frac{F_{CO\ J}}{Jy\ km/s}\right)(R_{J1})(1+z)^{-3}\left(\frac{\lambda_{obs\ J}}{mm}\right)^2\left(\frac{D_L}{Gpc}\right)^2 \quad (1).$$

In general the conversion factor α depends on the average cloud density $<n(H_2)>$, on the equivalent Rayleigh-Jeans brightness temperature $T_{R\ J}$ of the CO transition J→J-1, and on metallicity $Z$ (see Genzel et al. 2012 and Bolatto, Wolfire & Leroy 2012 for more detailed discussions),

$$\alpha_{CO\ J} = h\left(\frac{\left(<n(H_2)>\right)^{1/2}}{T_{R\ J}}\right)g(Z) \quad (2).$$

In the Milky Way and nearby star forming galaxies with near solar metallicity, as well as in dense star forming clumps of lower mass, lower metallicity galaxies, the empirical CO 1-0 conversion factors determined with dynamical, dust and γ-ray calibrations are broadly consistent with a single value of $\alpha_G$=4.36 ± 0.9 $M_\odot$/(K km/s pc$^2$), equivalent to $X_{CO}=N(H_2)/(T_{RJ=1}\Delta v)$= 2x10$^{20}$ (cm$^{-2}$/(K km/s), Strong & Mattox 1996, Dame, Hartmann & Thaddeus 2001, Grenier, Casandijan & Terrier 2005, Bolatto et al. 2008, Leroy et al. 2011, Abdo et al. 2010, Ostriker, McKee and Leroy 2010, Bolatto et al. 2012).

For the z~1-2.5 SFGs in this paper this 'Galactic' conversion factor is likely appropriate as well, since the CO emission in these systems, as in z~0 disk galaxies, probably arises in virialized giant molecular cloud systems (GMCs) with mean gas densities of $<n(H_2)> \sim 10^{2...3}$ cm$^{-3}$ (Dannerbauer et al. 2009, Combes et al. in prep., Daddi et al. 2010a; see also Magdis et al. 2012) similar to those in the Milky Way. Dust temperatures in the high-z and low-z main-sequence galaxy populations also are very similar ($T_{dust}$~30±7 K: Hwang et al.2010, Elbaz et al 2011).

For galaxies of gas phase metallicity less than solar, the conversion factor increases with $Z^{-1...-2}$ as the result of an increasing fraction of the molecular gas that is photo-dissociated in CO (Leroy et al. 2011, Genzel et al. 2012) relative to H$_2$.

To convert the CO 3-2 luminosity to an equivalent CO 1-0 luminosity we apply a correction factor of $R_{13}=L'_{CO\ 1-0}/L'_{CO\ 3-2}$~2 to correct for the lower Rayleigh-Jeans brightness temperature of the 3-2 transition relative to 1-0, caused by combination of the Planck correction (for a finite rotational temperature) and a sub-thermal population in the upper rotational levels. We take $R_{13}$=2 motivated by recent CO ladder observations in high-z SFGs (Weiss et al. 2007, Dannerbauer et al. 2009, Ivison et al. 2011, Riechers et al. 2010, Combes et al. in prep., Bauermeister et al. 2012). We then compute baryonic gas fractions from $f_{gas}=M_{mol\ gas}/(M_{mol\ gas} + M_*)$.

Obviously the quoted gas masses and depletion time scales in this paper scale are proportional to the conversion factor, and can thus be converted easily for different choices of $\alpha$. For the SFGs at z~1-3 gas fraction approximately scale with $(\alpha/4.36)^{0.6}$.

## *2.5 Kinematic classification*

We searched the CO data cubes for evidence of spatially resolved, monotonic velocity gradients that are the tell-tale signs of coherent orbital motion, and in particular of rotation. In all 9 sets with sub-arcsecond resolution data we detected such gradients, and in most of these the classification as a rotating disk is fairly obvious (section 3.1). Perhaps surprisingly at first, we also detected velocity gradients in the compact configuration data of 21 other SFGs, by exploiting the fact that relative centroids at different velocities can be measured much more accurately than the beam size if the signal to noise ratio is sufficiently high. Final classifications for each galaxy (as disk, merger, compact etc.) were obtained by combining the kinematic properties from the CO data, with the morphological information from the HST imaging that was available for all z~1.2 SFGs (see section 3.1).

To estimate the characteristic circular velocities of the SFGs we take the isotropic virial estimate ($v_c = \sqrt{\left(3/8\ln 2\right)}\Delta v_{FWHM}$) for unresolved galaxies without a velocity gradient, and $v_c = 1.3 \times \frac{\Delta v_{blue-red}}{2\sin(i)}$ if a velocity gradient ($\Delta v_{blue-red}$) indicative of rotation is detected in a galaxy with an inclination, $i$ (estimated from the morphological aspect ratio on the HST/Hα images; see Förster Schreiber et al. 2009). The factor 1.3 is an empirical calibration determined from disk models with source sizes, rotation velocities, spatial and spectral resolutions comparable to that of our data (Förster Schreiber et al. 2009).

## *2.6 Stellar masses, star formation rates and effective radii*

We derive stellar masses over the full redshift range using a consistent methodology based on SED modeling with assumptions regarding star formation histories and dust attenuation that are standard in the literature (e.g. Erb et al. 2006, Förster Schreiber et al. 2009, Wuyts et al. 2011b), all based on a Chabrier (2003)

IMF. For the z~2.2 sample these masses are thus directly comparable to those published in Erb et al. (2006), Förster Schreiber et al. (2009, and in prep.) and Law et al. (2009). Typical systematic uncertainties are ±0.13 dex.

Our adopted star formation rates are based on the sum of the observed UV- and IR-luminosities, or an extinction corrected Hα luminosity. For the z~1.2 sample we used the methods of Wuyts et al. (2011b) to combine the observed UV luminosities and Spitzer 24μm luminosities that are extrapolated with recent Herschel PACS calibrations to total infrared luminosities. For the latter we applied a single, luminosity independent FIR SED (Wuyts et al. 2008) to correct to the total far-infrared luminosity, as motivated by recent PACS-Herschel observations (Elbaz et al. 2010, 2011, Nordon et al. 2010, 2012). Star formation rates with a Chabrier IMF are then: $SFR_{UV+IR} = 1.087 \times 10^{-10} [L_{IR} + 3.3\, \nu L\nu(2800\text{Å})]$ (Wuyts et al 2011b).

In the case of Hα-based star formation rates ($SFR=L(H\alpha)_0/2.1\text{e}41$ erg/s, Kennicutt 1998b), we apply the additional Calzetti (2001) upward correction by a factor of 2.3 of the nebular extinction relative to that of the nearby stellar continuum ($A(H\alpha)=7.5\, E(B-V)$). Several authors (e.g. Förster Schreiber et al. 2009, Mancini et al. 2011, Wuyts et al. 2011b) find that this additional correction (typically resulting in a factor of ~2 upward correction of star formation rates) is preferable at z~2 (but see Erb et al. 2006, Reddy et al. 2010, 2012). Typical systematic uncertainties are ±0.15 dex.

Effective (=half-light) radii $R_{1/2}$ values are derived from Sersic fits to the HST-ACS and/or WfC3 CANDELS data (Grogin et al. 2011) for the EGS sample SFGs. Wherever we had both I-band (~rest-frame B-band) and H-band (~rest-frame R-band) data we averaged the estimates. In the case of the z~2.2 SFGs we adopt the Hα-sizes quoted by Erb et al. (2006) and Förster Schreiber et al. (2009). Typical uncertainties are ±0.1 dex.

# 3. Results

The derived galaxy properties of the observed z~1.2 and 2.2 SFGs are listed in Table 2. Our base sample has 38 detections at z=1-1.5 and 14 detections at z=2-2.5 (all at > 3σ), and 6 non-detections at z=2-2.5. In our final PHIBSS sample in Table 2 we also include the 6 CO J=2-1 detections of the z~1.5 BzK sample of Daddi et al. (2010a, labeled as 'BzK….'), and the 6 significant CO J=2-1 detections of the z~1 Herschel-PACS selected sample of Magnelli et al. (2012b, labeled 'PEPJ…'). These samples cover the same stellar mass-star formation range as our base sample. Finally we also include in Table 2 the CO detections of three somewhat lower mass ($M_*$=5-30 x$10^9$ $M_\odot$), strongly lensed SFGs between z~2.3 and 3.1 (cB58: z=2.7, Baker et al. 2004, 'cosmic eye": z=3.1, Coppin et al. 2007, 'eyelash': z=2.3, Swinbank et al. 2010, Danielson et al. 2011). The final PHIBSS sample thus has 50 CO detections at z=1-1.5 (with one SFG at z=1.6) and 17 detections at z=2-3. PHIBSS approximately triples the number of detections of CO emission in main-sequence galaxies at the peak of cosmic star formation.

Figure 1 shows the location of the PHIBSS galaxies in the stellar mass–star formation and stellar mass-specific star formation planes, compared to the two parent samples from zCOSMOS (z=1.5-2.5 BzK, McCracken et al. 2010, Mancini et al. 2011) and GOODS-S/CANDELS (Wuyts, priv.comm.). Figures 2, 3, 4 and 5 display the galaxy integrated CO 3-2 spectra, together with available HST images of the detections of our base sample. In the following sections we discuss the basic properties and inferences made from these measurements.

## 3.1 Kinematic and structural properties of SFGs: The majority are disks

In the following we assign simple kinematic classifications to the SFGs of our sample, from the combination of the derivable CO kinematic information and the morphological information from the HST images.

We classify an SFG as a 'quality A' rotating disk if it exhibits both an extended, disk-like HST morphology, and a significant, spatially resolved CO velocity gradient. 24/38 (63%) of our z~1-1.5 sample are quality A disks. For galaxies with disk-like HST morphologies but no detected velocity gradients (4/38), we classify the disks as quality B, for a total of 28/38 (74%) quality A+B disks at z~1-1.5. For 80% of these A or B quality disks we find a Sersic index between 0 and 2 from fitting I, J and/or H-band HST brightness distributions with GALFIT (Peng et al. 2002). In 3 cases we find wide-separation interacting disks, the remainder (5) are amorphous or compact systems, or obvious mergers (2). For the 6 SFGs of Daddi et al. (2010a,b) a similar analysis yields 3 quality A disks, plus 2 quality B disks.

For the z~2.2 sample the situation is analogous. We have sub-arcsecond resolution Hα integral field data sets from VLT SINFONI for 8 SFGs (from Genzel et al. 2008, 2011, Förster Schreiber et al. 2009, and in prep.) and spatially resolved Hα velocity gradients from Keck NIRSPEC slit spectroscopy for 2 (Erb et al. 2006). For 9 of the 20 z~2.2 SFGs there are HST NICMOS images from Förster Schreiber et al. (2011, and in prep) or HST/WFC3 images from Law et al. (2012a) and Lang et al in prep. The combination of the kinematic and morphological information yields (for the 13 cases where at least one is available) 7 (54%) quality A disks, 1 quality B disk, 2 compact SFGs, and 3 interacting or merging galaxies.

Combining all the samples described above (57 SFGs between z~1-2.5), we find that 34 (60%) are quality A disks, 41 (72%) quality A+B disks, 8 (14%) are compact/amorphous and 8 (14%) are interacting and/or merging. More than two thirds of the z~1.2 and 2.2 main-sequence SFGs discussed in this paper are (large) rotating disks, in excellent agreement with the morphological analysis of the surface brightness distributions of much larger galaxy samples drawn from HST imaging surveys (Wuyts et al. 2011a).

We caution that these are first-order classifications, and may be too simplistic, especially in the ability to identify minor mergers. Rest-frame UV- and B-band HST images of many z~1-1.5 SFGs and almost all z~2-3 SFGs are clumpy and asymmetric, even if rotational support dominates (Figures 2-5, Elmegreen et al. 2005, Wuyts et al. 2012). A number of the HST images in Figures 2-5 also exhibit clumps or filamentary structures outside the main body of the galaxies. It is a priori not clear whether these are indications of ongoing minor mergers or interactions, or whether they are clumpy star forming regions at large radii, or even whether they are unrelated to the main galaxy. On average, rest-frame optical HST images and inferred stellar surface density distributions are typically much smoother, however, for the z=1-2.5 main-sequence population, suggesting that most clumps are relatively short-lived star formation regions in a globally unstable, gas rich disk (Elmegreen et al. 2007, Wuyts et al. 2012).

### 3.1.1 CO and UV/optical sizes are similar

Our sub-arcsecond observations for 9 SFGs map out the intragalactic spatial distribution and kinematics of CO relative to the UV/optical stellar distributions from the HST. Figure 6 presents overlays of the CO integrated flux distributions (white contours) on the HST I- or H-band images (rest-frame B- or R-band). With the exception of the interacting galaxy pair (EGS12011767) the CO emission qualitatively traces the stellar light distribution quite well. For an additional 8 z=1-1.5 SFGs observed in C/D configuration we were able to derive significant CO sizes from circular Gaussian fits in the UV plane. In Figure 7 we compare the HST I/H-band half-light radii with the radii inferred for the CO 3-2 emission, either by n=1 Sersic fits in the image plane (for the well resolved sources), from circular Gaussian fits in the U-V-plane, or both. We also include 4 z~1.5 SFGs, for which Daddi et al. (2010a) were able to derive CO radii. In two cases we compare the CO sizes with Hα sizes and with the fitted stellar mass distributions, inferred from two band J/H-HST imagery using the methods of Förster Schreiber et al. (2011) and Lang et al, in prep.

All these estimates of $R_{1/2}$ carry substantial uncertainties, either because of the influence of extinction, clumpy individual star forming regions and/or the presence of central bulges (in the UV/optical stellar light), or because of the relatively poor spatial resolution and dynamic range of the CO data. We have not removed the light from the nuclear bulges in estimating the optical sizes, which typically affect most prominently the H-band data and the stellar mass maps, as a result of which the H-band radii and half-mass radii are somewhat below the dotted diagonal line in Figure 7. A linear fit to the SFGs with CO and I-band radii yields $R_{1/2}(CO) = 1.02$ (±0.06) $R_{1/2}$ (rest B-band).

Molecular gas and optical light distributions have comparable overall radial scales, in agreement with the same finding in z~0 disk galaxies (e.g. Young et al. 1995, Regan et al. 2001, Leroy et al. 2008). In the three cases with estimated intrinsic stellar mass distributions (BX610, EGS13018632 and EGS13011166), the half-mass radius is smaller than that of the gas and current star formation (as traced by CO and Hα). This is consistent with Nelson et al. (2012) who found R(Hα)/R(stars)~1.3 for a sample of 57 z~1 SFGs in the 3D-HST survey. In the case of BX610 the Hα and J-band distributions appear to come from an inclined ring rotating around the central gas rich region, within which a red bulge is embedded. The more compact stellar mass distributions thus are probably the result of significant older bulges embedded in younger star forming disks.

### 3.1.2 Ratio of rotation to random motions: z>1 SFGs are turbulent

A generic characteristic of all z>1 SFGs is an extended 'floor' of large local velocity dispersion in their Hα velocity fields, $<\sigma_0>$ ~30-100 km/s (after removal of beam-smeared orbital motion and instrumental resolution), so that the ratio of rotational motion to velocity dispersion is typically ~5 (for the same stellar mass range as for the sample in this paper, $logM_*$~10.4…11.5; Genzel et al. 2008, 2011, Förster Schreiber et al.2009, in prep., Law et al. 2009, Epinat et al. 2009, 2012, Wisnioski et al. 2011, 2012, Law et al. 2012b). High-z ionized gas disks are turbulent and geometrically thick. The UV-light distributions of edge-on 'chain' galaxies also indicate z-scale heights (~1 kpc) comparable to those of the ionized gas (Elmegreen & Elmegreen 2006). An important question then is whether the large turbulence is a characteristic of the entire gas layer, or only of the ionized gas and stellar light.

In the lensed 'eyelash' system (J2135-0102, z=2.33), Swinbank et al. (2011) found that the spatially resolved molecular velocity dispersion is 50±10 km/s, comparable to that of the ionized gas. The high resolution data sets of our sample allow for the first time to make such measurements for a larger set of galaxies, with the same methods as, for instance, applied in Genzel et al. (2008, 2011) and Davies et al. (2011). Briefly, Gaussian profiles are fitted to each spatial pixel. For a spatially resolved, rotating disk with a constant floor of velocity dispersion, the effect of beam-smeared rotation is minimized along the line of nodes in the outer parts of the galaxy.

The average velocity dispersion measured in this region, after correction for the instrumental resolution, is a good estimate of the intrinsic local velocity dispersion (Davies et al. 2011), although for the ~4-8 kpc linear resolution of our CO maps there may be some residual beam smearing effects in the velocity dispersion estimates. Table 3 summarizes the results for these sources. The full analysis will be presented in Combes et al. (in preparation) and Freundlich et al. (in preparation).

To first order the velocity dispersions and $v_c/\sigma_0$ ratios in CO and those in Hα agree. There may be a slight trend for the average values of $v_{rot}/\sigma_0$ to be somewhat larger for the molecular gas than for the ionized gas (~7 compared to ~5). The difference is marginally significant, given the dispersions of the respective distributions (each ~±1.7). Rotating disks at z~0 have $v_{rot}/\sigma_0$~10-20 (Dib, Bell & Burkert 2006). Large random motions indeed appear to be an intrinsic property of the entire ISM of high-z SFGs.

## 3.2 Specific star formation scales with gas fraction

Of the various scaling relations between gas and galaxy properties that we will discuss in this paper the relationship between specific star formation rate and gas fraction, defined as $M_{mol\ gas}/(M_{mol\ gas} + M_*)$, (or the ratio $M_{mol\ gas}/M_*$) has the lowest scatter. This is shown in the left and right panels of Figure 8. Specific star formation rate $sSFR=SFR/M_*$ and molecular gas fraction $f_{mol\ gas}$ are related through

$$f_{mol\ gas} = \frac{M_{mol\ gas}}{M_{mol\ gas} + M_*} = \frac{1}{\left(1 + \left[sSFR \times t_{depl}\right]^{-1}\right)} \qquad (3),$$

where $t_{depl}=M_{mol\ gas}/SFR$. The z=1-1.5 data points (filled black circles) are very well fit by equation (3) with a constant depletion time scale ($t_{depl}$ ~7x10$^8$ yr). The fewer

z=2-2.5 points (crossed red squares) are consistent with the same relation but the scatter is larger, perhaps in part because of the metallicity dependence of the conversion factor for 5 of the 15 data points (Genzel et al. 2012). The strong trend is seen even more clearly in the right panel of Figure 8, where we plot the ratio of molecular gas mass to stellar mass as a function of *sSFR*. The best fit relation has a slope of 0.7 (±0.2) and a dispersion of ±0.24 dex around this fit. Formally the systematic uncertainties in *sSFR* and gas to stellar fractions are ±0.2dex and ±0.25dex (see discussion in Erb et al. 2006, Förster Schreiber et al. 2009, Genzel et al. 2010). These values are comparable to the residual scatter in Figure 8, suggesting that the intrinsic correlation between *sSFR* and gas fraction is even tighter. The offset from the mean main-sequence for z~1-2.5 SFGs thus is largely controlled by how gas rich a given galaxy is. Magdis et al. (2012) reach a similar conclusion based on a completely independent method of determining dust masses from Herschel far-infrared SEDs, and inferring gas masses and fractions from the mass-metallicity relation and the relation of Leroy et al (2011) between metallicity and gas-to-dust ratio in z~0 SFGs.

In addition galaxies above the main-sequence also have smaller depletion time scales (Kennicutt & Evans 2012, Saintonge et al. 2012). This is demonstrated most clearly by observations of z~0-4 galaxies far above (factor 5-10) the main-sequence line. Most objects in this region appear to be dominated by short-lived starburst events with ~5-50 times smaller depletion time scales than in the main-sequence SFGs (Daddi et al. 2010b, Genzel et al. 2010, Magnelli et al. 2012a). Many of the most luminous submillimeter galaxies (SMGs) at z~1-4 also appear to be representatives of this 'starburst' or 'merger' mode (Engel et al 2010, Tacconi et al 2008). Corrected for their ~4 times smaller CO-conversion factors, these extreme SMGs have molecular gas fractions comparable to their main-sequence cousins (Greve et al. 2005, Tacconi et al. 2008, Engel et al. 2010, Bothwell et al. 2012). In the z~0 COLDGASS survey SFGs near the main sequence may also exhibit an inverse correlation between depletion time scale and *sSFR*, although this correlation is weaker than the trend with gas fraction (Saintonge et al. 2011b, 2012).

### *3.3 The KS- relation at z=1-3 is near-linear*

The much larger sample of z~1-3 SFGs in this paper (as compared to that in Genzel et al. 2010 and Daddi et al. 2010b) allows us to revisit the discussion of the molecular KS-relation (Kennicutt 1998a, Kennicutt & Evans 2012). Figure 12 shows the results. The 50 CO detections are now sufficient to determine independently the slope of the z=1-1.5 relation, and not just the intercept, as in the 2010 papers. It is apparent that this data set is well fit by a linear KS-relation ($\Sigma_{star\,form}=\Sigma_{mol\,gas}/(0.7 (\pm0.05)$ Gyr)). Within the uncertainties, the z~2-3 SFGs lie on the same relation. The dispersion around a linear relation is $\pm0.25$ dex, which is comparable to the total measurement uncertainties, again suggesting that the intrinsic relation is quite tight. A weighted bivariate fit without restricting the slope to all z=1--3 SFGs with {12 + log(O/H)>8.6}, yields $log\Sigma_{starform}=-3.01(\pm0.32)+1.05(\pm0.11)*log\Sigma_{molgas}$. Fitting only the z=1-1.5 data, or fitting with a standard least squares technique, gives similar results with inferred slopes ranging between 0.8 and 1.15, each with a formal 1$\sigma$ uncertainty of about $\pm0.15$. A corresponding bivariate fit to all 93 z=0 near-main sequence SFGs in Figure 9 yields $log\Sigma_{starform}=-3.5(\pm0.1) +1.1(\pm0.05)*log\Sigma_{molgas}$.

An intrinsic slope significantly steeper than unity would be required if the conversion factor α decreases systematically with $\Sigma_{mol\,gas.}$ Such a trend is plausible if larger surface densities correlate with high temperature (see the discussion in Appendix A and Figure 10 of Tacconi et al. 2008, and in Kennicutt & Evans 2012).

For instance, if the conversion factor changes from the Milky Way value ($\alpha$=4.36) at $log\Sigma_{mol\,gas}$=2 to the 'ULIRG' value ($\alpha$~1) at $log\Sigma_{mol\,gas}$=4, the resulting slope of the relation in Figure 9 would be N=1.2.

In the case of the massive, clumpy rotating disk EGS13011166 (z=1.53, Figure 4 upper right, left in middle row of Figure 6), we have data cubes of CO 3-2 and H$\alpha$ (from LUCI on the LBT) at a similar resolution of ~0.7" FWHM (PHIBSS team, in preparation). The agreement in spatial distribution, velocity field and velocity dispersion field of the two tracers is remarkable. These measurements allow for the first time a determination of the z>1 intragalactic KS-relation at a resolution of ~6 kpc. The cyan squares in Figure 9 show the results. Here we have set the absolute scale of the H$\alpha$ data points by assuming that the H$\alpha$-derived star formation rate is the same as that obtained from our standard SED technique described in section 2.6. The inferred slope of the spatially resolved KS-relation in EGS13011166 is 0.95 ($\pm$0.05), and the depletion time scale 0.77 ($\pm$0.1) Gyr.

A comparison of our spatially resolved CO data in EGS12007881, EGS13003805, EGS13004291 and EGS13019128 with the [OII] line emission in 1-1.5" DEEP2 slit spectra (as a tracer of star formation activity) yields a linear KS-relation as well, and an average depletion time scale of 1.05 ($\pm$0.4) Gyrs (Freundlich et al. 2013).

### 3.3.1 Impact of gas excitation and observed J-level on the inferred z>1 KS-relation

Several recent papers have commented on the possible impact of varying excitation in different galaxies on the inferred KS-relation (Ivison et al. 2011, Narayanan et al. 2011a). Narayanan, Cox & Hernquist (2008) and Narayanan et al. (2011a) have found from simulations that the KS-slope decreases systematically with the excitation of the tracer, because higher excitation lines (such as the J=3-2 CO line) are insufficiently excited in lower density and lower column density gas. In these simulations the low excitation CO 1-0 emission faithfully tracks the intrinsic KS-slope, while the 3-2 line has a shallower slope, by $\Delta N \sim 0.2$-$0.5$. To check on this possibility, we plot in the right panel of Figure 10 the correlation between CO 3-2 luminosity and star formation rate in z~0 SFGs, LIRGs and ULIRGs (from Wilson et al. 2012, Iono et al. 2009 and Papadopoulos et al.2010), in z~1-3 SFGs from PHIBSS and in z~1-4 SMGs (Bothwell et al. 2012). With the exception of the ULIRG and LIRG mergers, all near-main sequence SFGs and the SMGs in this diagram are compatible with the same near-linear slope as found in Figure 10. It is thus unlikely that this slope is an artifact of excitation variations.

### 3.3.2 Impact of galaxy interactions and dense radiation fields on the KS-relation

The next question is how sensitive the molecular KS- relation is to the perturbation of the galactic ISM by interactions (minor and major mergers) and the density of the radiation field. We explore this issue in the left panel of Figure 10. In order to separate out the issue of the CO-conversion factor, we adopt the same Milky Way conversion factor for all galaxies in this diagram. The horizontal arrow denotes how galaxies move if instead of the Milky Way value of $\alpha$=4.36, a 'ULIRG' conversion factor of $\alpha \sim 1$ is chosen.

The plot shows that inferences about the slope of the gas-star formation relation and the depletion time are relatively insensitive to the dynamical state and radiation field for z~0 SFGs over a wide range of properties, unless one reaches conditions of major dissipative mergers. Moderate z~0 starburst galaxies and isolated luminous infrared galaxies (LIRGs), such as M82, NGC253, NGC1614, NGC2146, or NGC7469 (located at $log\ \Sigma_{star\,form}$ ~ -0.5..+0.6), with radiation field densities $G$ ~100-1000 times that of the radiation field density in the solar neighborhood) are located very close to the relation of normal disks ($log\ \Sigma_{star\,form}$ ~ -2.5…-1), with similar depletion time scales (see also Saintonge et al. 2012). Galaxies with strong bars are also on the same relation (Saintonge et al. 2012). Galaxies involved in interactions and mergers are above the 'disk' line but for a constant conversion factor only strongly disturbed pairs or advanced mergers in LIRG/ULIRG systems are high above the 'disk' relation and have much smaller depletion time scales (Figure 9, Saintonge et al. 2012). The effect of the likely smaller conversion factors in these systems exacerbates these differences, as shown by Genzel et al. (2010) and Daddi et al. (2010b). Such systems are 0.7-1 dex above the main-sequence.

Given that most of the z~1-3 PHIBSS SFGs are more like local starbursts/isolated LIRGs in terms of their radiation fields and dynamic state, the adoption of a constant depletion time scale across most of the z~1-3 main-sequence population thus seems plausible. Similar to the situation at z~0, z~1-3 galaxies high above the main-sequence, such as many SMGs, are often in major mergers (Tacconi et al. 2006, 2008, Engel et al. 2010), and have much smaller depletion time scales.

In summary, the galaxy integrated data of z~1-3 near-main sequence SFGs favor a near-linear molecular gas–star formation relation, similar to the findings at z~0 (Bigiel et al. 2008, 2011. We caution that the systematic uncertainties due to conversion factor variations, different sample selections and analysis methods are substantial. As discussed in Genzel et al. (2010), the systematic uncertainties in slope determinations are probably at least ±0.2 dex. For simplicity, however, we will now adopt a linear relation, N=1, with a constant depletion time scale of 0.7 Gyr for the

z=1-3 SFG population, consistent with the independent results of Magdis et al. (2012).

### 3.3.3 Need for gas replenishment

The right panel of Figure 9 compares the depletion time distributions of our z=1-1.5 PHIBSS sample with a z~0 COLDGASS comparison sample, with the same mass range and location relative to the z~0 main-sequence line. Both distributions have a comparable dispersion (±0.24 in *log $t_{depl}$*). Their zero points differ by 0.24 in the log (1.7 in depletion time), such that <$t_{depl}$> (COLDGASS comparison sample) =1.24 (±0.06) Gyr. Note the difference to the larger depletion time scale of the entire z~0 main-sequence population averaged over a larger mass and star formation range (1.5±0.5 Gyrs between *logM*$_*$=10 and 11.3, Bigiel et al. 2008, 2011, Leroy et al. 2008, Saintonge et al. 2011b).

The depletion times in both redshift ranges are significantly smaller than the Hubble time, $t_{depl}/t_{Hubble}(z)$~0.1-0.2. Along with the finding of large star formation duty cycles in main-sequence SFGs (30-70%: Noeske et al. 2007, Daddi et al. 2007, Reddy et al. 2005), the short depletion timescales indicate that there must be replenishment of the gas reservoirs from the halos or circum-galactic regions in most SFGs at the peak of the cosmic star formation activity (see also Tacconi et al. 2010, Genzel et al. 2010 and Daddi et al. 2010b). The combination of the observed large duty cycles of the main sequence population and the near-linear KS-relation favoring a single depletion time suggests that the scatter in the stellar mass-star formation relation is largely dominated by fluctuations in the accretion rate, and not by the 'burstiness' of the star formation processes within galaxies. The depletion time scale at z~1.2 is about half that at z~0, implying a scaling as $\propto(1+z)^{-0.7...-1}$. This is somewhat shallower than what is expected if the depletion time scale is proportional to the Galactic dynamical time scale, $t_{dyn}(v_c)=R(z)/v_c \propto (1+z)^{-1.5}$ (Davé et al. 2011, 2012).

### *3.4 Large molecular gas fractions in z>1 SFGs*

Figure 11 shows the distribution of the *observed* molecular gas fractions at z~1.2 and z~2.2 determined from the PHIBSS data. The gas fraction distribution is broad with a scatter of $\sigma_f$=0.18 centered on $<f_{mol\,gas}>$=0.49±0.02 at z~1.2 and 0.47±0.05 at z~2.2, in broad agreement with the results of Tacconi et al. (2010) based on a 2.5 times smaller sample.

These numbers are subject to significant uncertainties and corrections. The absolute level of the derived molecular gas fractions obviously depends on the value of the adopted conversion factor, as we already discussed in section 2.4. For the z~1-1.5 sample a 'merger'-conversion factor ($\alpha$~1 instead of 4.36) may be more appropriate for two probable merger systems, EGS12004351 and EGS13004291. Omitting these two outliers from the sample does not change the median gas fraction.

The dependence of the conversion factor on metallicity should not affect the z=1-1.5 SFGs, since their metallicities, based on the stellar mass-metallicity relation, are all estimated to be >0.8 times the solar value. However, 6 of the 17 z~2-3 CO detections and 2 of the 6 non-detections listed in Table 2 are in this lower metallicity regime, where metallicities are estimated from the mass-metallicity relation or the [NII]/Hα emission line ratio estimator (see Genzel et al. 2012 for a full description). The z~2.2 gas fractions in these cases may well be lower limits, as indicated by the arrows in Figure 11.

### 3.4.1 Correction for incomplete coverage of the $M_*$- SFR plane

The most important effect is the extrapolation from the observed average gas fractions to the entire z~1-3 star formation main-sequence population. This effect arises primarily because the parameter space in the stellar mass- (specific) star formation rate plane below the main-sequence line is incompletely or not covered for the z~1-1.5 SFGs, given the 30 $M_\odot$ yr$^{-1}$ SFR lower limit selection cut of the PHIBSS sample (bottom left panel of Figure 1). The coverage is more complete for the z=2-2.5 SFGs (bottom right panel in Figure 1). A second correction arises because the coverage of observed sources above the star formation cut is close to homogeneous, rather than reflecting the true parent sample distribution, which is concentrated around the main-sequence line.

To correct for this incomplete sampling, we exploit that, to first order, the main sequence population is well-described by a constant gas depletion time scale of ~0.7 Gyr. If this constant depletion timescale holds for the entire main-sequence population at z~1-3, we can take the observed distribution of SFGs in the stellar mass – (specific) star formation plane in the parent imaging surveys to compute molecular gas fractions using equation 3. To do this we have taken the results from the zCOSMOS star forming (s)BzK z=1.5-2.5 survey (McCracken et al. 2010, Mancini et al. 2011; see upper two panels in Figure 1), the CANDELS GOODS-S sample (Wuyts et al. priv. comm.; see bottom two panels in Figure 1), and the NEWFIRM medium-band survey in the AEGIS and COSMOS fields (Whitaker et al. 2012). At redshift 2-3, we then find that the average gas fractions inferred from these surveys (within ±0.7 dex of the main-sequence line) are 0.45 to 0.5, similar to or slightly below the direct measurement shown in Figure 11. The inferred gas fractions would be larger if the gas depletion time scale at z~2-2.5 were smaller than 0.7 Gyr, as discussed in section 3.6.

At z~1-1.5 the incompleteness correction is larger, as expected. Correction for the homogeneous coverage above the star formation cutoff decreases gas fractions from 0.49 down to ~0.4, the incomplete coverage below 30 $M_\odot$ yr$^{-1}$ leads to a further decrease to about 0.3-0.35.

These extrapolated molecular gas fractions are about 4 to 6 times greater than for average z~0 main sequence SFGs for the same mass range ($f_{mol\,gas}$=0.08). That ratio is 4 when one matches the COLDGASS sample to the same mass range and relative location in the mass-star formation plane as at z~1-2.5 ($f_{mol\,gas}$= 0.12, Saintonge et al. 2011a).

### 3.4.2 Contribution of atomic hydrogen & photodissociated gas

In the local Universe the ratio of atomic to molecular gas masses in $logM_*$~10…11 SFGs is about 1.5-2 (Saintonge et al. 2011a), and the atomic gas dominates the mass of the cold ISM. The $H_2$/HI ratio in the ISM is strongly pressure/column density dependent and exceeds unity above a column density of ~10-15 $M_\odot$ pc$^{-2}$ (~1-1.5x10$^{21}$ cm$^{-2}$, Sternberg & Dalgarno 1995, Blitz & Rosolowsky 2006, Krumholz, McKee & Tumlinson 2009). Therefore the HI/$H_2$ ratio is expected to decrease strongly at high-z, as the mean $H_2$ column densities and ISM pressure are at least an order of magnitude greater than at low-z (e.g. Obreschkow & Rawlings 2009, Lagos et al. 2011, Fu et al. 2012). Since most of the high-z disks discussed in this paper have hydrogen columns >10$^{2.5}$ $M_\odot$ pc$^{-2}$ (Table 2), the molecular fractions of the disks should be very high.

Another issue is the fraction of $H_2$ gas that is photo-dissociated in CO. Wolfire, Hollenbach & McKee (2010) estimate that for solar metallicity, and for typical UV fields of $G$~1-100, densities and column densities in local and high-z SFGs, the mass fraction of this 'CO-dark' gas may be between 20 and 30%. This fraction increases strongly for lower metallicities, leading to a strong correction in the CO-conversion factor, as discussed above.

### *3.5 Gas fractions drop with increasing stellar mass*

Figure 12 summarizes the dependence of the observed z=1-1.5 molecular gas fractions as a function of stellar mass. We divide the sample of 50 SFGs into three mass bins ($2.5 \times 10^{10} < M_* < 5 \times 10^{10}$, $5 \times 10^{10} < M_* < 10^{11}$, $10^{11} < M_* < 3 \times 10^{11}$ $M_\odot$), such that there are a comparable number of galaxies in each bin. The scatter in each bin is substantial but the median gas fractions show a significant (4 – 5 σ) drop from 0.58 to 0.37, over about an order of magnitude in stellar mass. The inferred drop in gas fraction with stellar mass is in good agreement with the independent dust-mass method of Magdis et al (2012).

As discussed in the last section, the observed gas fractions need to be corrected for the incomplete coverage of the intrinsic underlying distribution of star forming galaxies in the stellar mass- (specific) star formation plane. The green shaded region in Figure 12 shows the distribution of inferred, corrected gas fractions obtained by applying equation (3), with $t_{depl}$=0.7 Gyr, to recent imaging UV/optical imaging surveys of z=1-2.5 SFGs that follow a *sSFR* – stellar mass relation as: *<sSFR>*= a*$(M_*/6.6 \times 10^{10}$ $M_\odot)^p \times ((1+z)/2.2)^q$, with a=0.68 (±0.1) Gyr$^{-1}$, p=-0.4(-0.1,+0.3), q=2.8(-0.5,+0.1) (Noeske et al. 2007, Daddi et al. 2007, Panella et al. 2009, Rodighiero et al. 2010, Bouché et al. 2010, Karim et al. 2011, Whitaker et al. 2012, Salmi et al. 2012, Wuyts et al. 2012).

If the assumption of a mass independent depletion time scale is valid, and depending on the scaling with mass of the *M*-sSFR* relation in this range, SFGs with smaller stellar masses than those sampled by the PHIBSS survey should even have larger gas fractions (Figure 12). In the COLDGASS sample Saintonge et al. (2011b) find $d\,(\log t_{depl})/\,d\,\log M_* = -0.31$. If a similar scaling also holds at z~1-1.5, the inferred gas fractions in Figure 12 at the broad peak at $\log M_*$=9.7-10 may come down by ~15%.

The right panel of Figure 12 indicates that although the gas fractions at z~1-2.5 (for $SFR$>30 M$_\odot$ yr$^{-1}$) are on average 4 times larger than for z~0 main-sequence SFGs in the same mass range (Saintonge et al. 2011a,b) the relative mass dependence (normalized to the gas fractions at $5\times10^{10}$ M$_\odot$) is remarkably similar in both redshift ranges. Assuming that this similarity holds also at larger stellar masses than those sampled in PHIBSS ($\log M_*$>11.3), the COLDGASS data suggest that the gas fractions continue to decrease rapidly in this mass range, in part because of the larger fraction of red and dead massive galaxies located below the main sequence. Note that the analysis of the COLDGASS sample suggests that gas fractions are more strongly correlated with stellar surface density or concentration, than with stellar mass (Saintonge et al. 2011b, 2012, Kauffmann et al. 2012). Future observations are required to test this hypothesis at z>1.

Recent simulations suggest that the relative stellar mass dependence of gas fractions on the star formation main sequence is controlled by the mass dependence of feedback and gas accretion efficiency onto the galaxy. In the mass range sampled by PHIBSS and COLDGASS the dominant effect may be the former (Davé et al. 2011, 2012). This is demonstrated in Figure 13 by plotting the predictions of different wind models from the work of Davé et al. (2011).

This analysis shows that the mass dependence of the incompleteness corrected, average gas fractions of z~0 and z~1-2.5 SFGs in the stellar mass range $10^{10...11.5}$ M$_\odot$ is remarkably well matched by momentum driven wind models proposed by Oppenheimer & Davé (2006, 2008), including the extrapolation of the gas fractions to lower masses shown in Figures 12 and 13. These models have winds with high mass loading ($\dot{M}_{wind}/SFR \sim 1$), consistent with recent observations (Erb 2008, Genzel et al. 2011, Newman et al. 2012). Perhaps surprising at first, these models with powerful winds predict that z~1-2 massive galaxies are gas rich (Genel et al. 2012), consistent with the observations, mainly because the efficient feedback prevents more complete conversion of gas to stars at yet higher redshifts. Models with no winds, constant winds, or weaker supernova-only feedback (Fu et al. 2012) do not provide a good match to the observed gas fractions at the high mass end.

The inferred factor ~1.4 (±0.4) drop in the average incompleteness corrected gas fractions from z=2-2.5 to z=1-1.5 is also consistent with published theoretical work. Simulations and semi-analytic results predict ratios of $f_{gas}$(z~2.2)/$f_{gas}$(1.2) ranging between ~1 and 2 (Ocvirk et al. 2008, Guo et al. 2010, Dutton, van den Bosch & Dekel 2010, Davé et al. 2011, Genel et al. 2012). It is important to keep in mind that if the cosmic accretion rate were the only factor driving the gas fractions, a drop of about a factor 2.3 between z~2.2 and 1.2 would be expected ($dM_{in}/dt \sim (1+z)^{2.2}$, Neistein & Dekel 2008, Genel et al. 2008). Flatter redshift dependences and simultaneously larger gas fractions at z<2 are generically achieved in models with stronger feedback (Figure 13), and in particular by models with efficient late-time recycling of gas ejected at earlier epochs ('galactic fountains'), such as in the work of Davé et al. (2011).

## *3.6 The cosmic evolution of specific star formation rate is mainly due to the evolution of gas fractions*

In the equilibrium (or 'bathtub') models of galaxy evolution (e.g. Bouché et al. 2010; Davé et al. 2011, 2012) the gas masses and gas fractions of main-sequence SFGs are set by the balance of gas accretion from the halo (including returning gas that was ejected at earlier epochs), star formation, gas return from stellar evolution and ejection of gas by stellar feedback. The star formation rate as a function of cosmic epoch is then tied to the available gas reservoir. In section 3.3 we showed that the relation between molecular gas mass and star formation rate is near-linear, with a well-defined, slowly variable depletion time scale for all near main-sequence SFGs at a given redshift. For simplicity we assume that the depletion time scale scales with redshift as $t_{depl} \sim 1.5 \times (1+z)^{-1}$ Gyr. With this assumption equation (3) can be inverted to solve for $sSFR$ ($f_{gas}$, $t_{depl}$). In Figure 14 we plot the average values of $sSFR$ inferred in this way as a function of redshift for the z=1-1.5 and z=2-2.5 redshift ranges, for z~0 from COLDGASS for the same mass range (Saintonge et al. 2011a,b), for 5 galaxies at z=0.39 from Geach et al. (2011) and 4 galaxies at z~0.3 from Bauermeister et al. (2012). We compare these gas based estimates of $sSFR$ in the equilibrium model with the direct observations from various photometric surveys, as compiled most recently in Weinmann et al. (2011), Sargent et al. (2010) and Wuyts et al. (2011a, and private communication). We refer the reader to those papers for the individual references to the data points shown.

The agreement of these two independent methods in the $sSFR$(z) evolution is impressive. It is thus plausible that the observed cosmic variations in specific star formation rates are indeed primarily controlled by the available molecular gas reservoirs. This plot also lends additional support to our assumption of a constant CO conversion factor at all redshifts for the main sequence SFG population.

# 4. Summary

We have surveyed the molecular gas properties in massive, main-sequence star forming galaxies (SFGs) near the cosmic star formation peak at redshifts from ~ 1 to 3. The IRAM Plateau de Bure high-z blue sequence survey (PHIBSS) provides for the first time sizeable samples of CO 3-2 line detections in two redshift slices selected by stellar mass (>2.5x10$^{10}$ M$_\odot$) and star formation rate (>30 M$_\odot$yr$^{-1}$). The lower redshift slice at z~1.2 contains 50 galaxies, including 12 SFGs from the literature. All targets in this redshift range have been detected in CO. The higher redshift slice at z~2.2 consists of 23 SFGs, 17 with detected (≥3σ) CO emission, including three lensed galaxies from the literature. PHIBSS triples the number of normal high-z SFGs with directly measured cold gas properties. Our main results are as follows:

- Conversion of the observed CO luminosities to molecular gas masses with a Milky Way conversion factor yields comparable average observed molecular gas fractions (including a correction for helium) of ~50% at z~1.2 and 2.2. Correction of these observed values for the incomplete coverage of the stellar mass – (specific) star formation plane in our observations lowers the z=1-1.5 and 2-3 gas fractions expected for the entire main-sequence population to about 0.33 and 0.47. This compares to 0.08 for z=0 SFGs selected in the same stellar mass range. Molecular gas fractions are stellar mass dependent. They drop by 60% between logM$_*$=10.5 and 11.3, consistent with the expectations from cosmological simulations with strong star formation feedback.
- Based on the comparison of HST morphologies and gas kinematics, between 50 and 75% of the z~1-2.5 SFGs are rotating spirals/disks (including disks with signs of modest perturbation, such as minor mergers), while ≤20% show evidence for major merging and strong interactions. The CO emission in the large disks comes from a region of size comparable to that of the UV/optical emission;

- For spatially well resolved SFGs we find that the ratio of rotational velocity to local velocity dispersion is ~7, consistent with but perhaps somewhat larger than dispersions in similar SFGs from Hα imaging spectroscopy. Large velocity dispersions are characteristic of the entire gas layer of high-z SFGs;

- The relation between galaxy-integrated molecular gas and star formation surface densities (the molecular 'Kennicutt-Schmidt' relation) is near-linear at all redshifts sampled so far, with a z~1-2 gas depletion time scale of ~0.7 Gyrs. In the five z~1-1.5 SFGs where we have spatially resolved CO and Hα/[OII] observations (PHIBSS team in prep., Freundlich et al. 2013), this linear dependence is confirmed on sub-galactic scales. Since near-main sequence SFGs have a high duty cycle of star formation, the relatively short depletion time scale requires semi-continuous replenishment on a local Hubble time scale. The CO observations support other evidence that the z~1-2 star formation main-sequence population consists largely of gas-rich disk galaxies in approximate equilibrium between incoming fresh gas from the halo, star formation in the disk and gas expulsion by galactic winds;

- Gas fractions correlate strongly with the specific star formation rate, *sSFR=SFR/M$_*$*, both at a given redshift as a function of star formation rate near the main sequence, and as a function of redshift. This suggests that at constant stellar mass, both the vertical location of a galaxy in the *M$_*$-SFR* plane, and the variation of specific star formation rate between z~0 and 2.5, are mainly driven by the available molecular gas reservoir. The combination of the observed large duty cycles of the main sequence population and the near-linear KS-relation favoring a single depletion time suggests that the scatter in the stellar mass-star formation relation is largely dominated by fluctuations in the accretion rate, and not by the burstiness of star formation processes within galaxies.

**Acknowledgements:** The observations presented here would not have been possible without the diligence and sensitive new generation receivers and WidEx backend from the IRAM staff – for this they have our highest admiration and thanks. We also thank the astronomers on duty and telescope operators for delivering consistently high quality data to our team. We are also grateful to Amber Bauermeister and Leo Blitz for sending us their data points in advance of publication, and to David Law for sending FITS images of several of the z~2 sources. ADB wishes to acknowledge partial support from a CAREER grant NSF-AST0955836, and from a Research Corporation for Science Advancement Cottrell Scholar award.

Figures

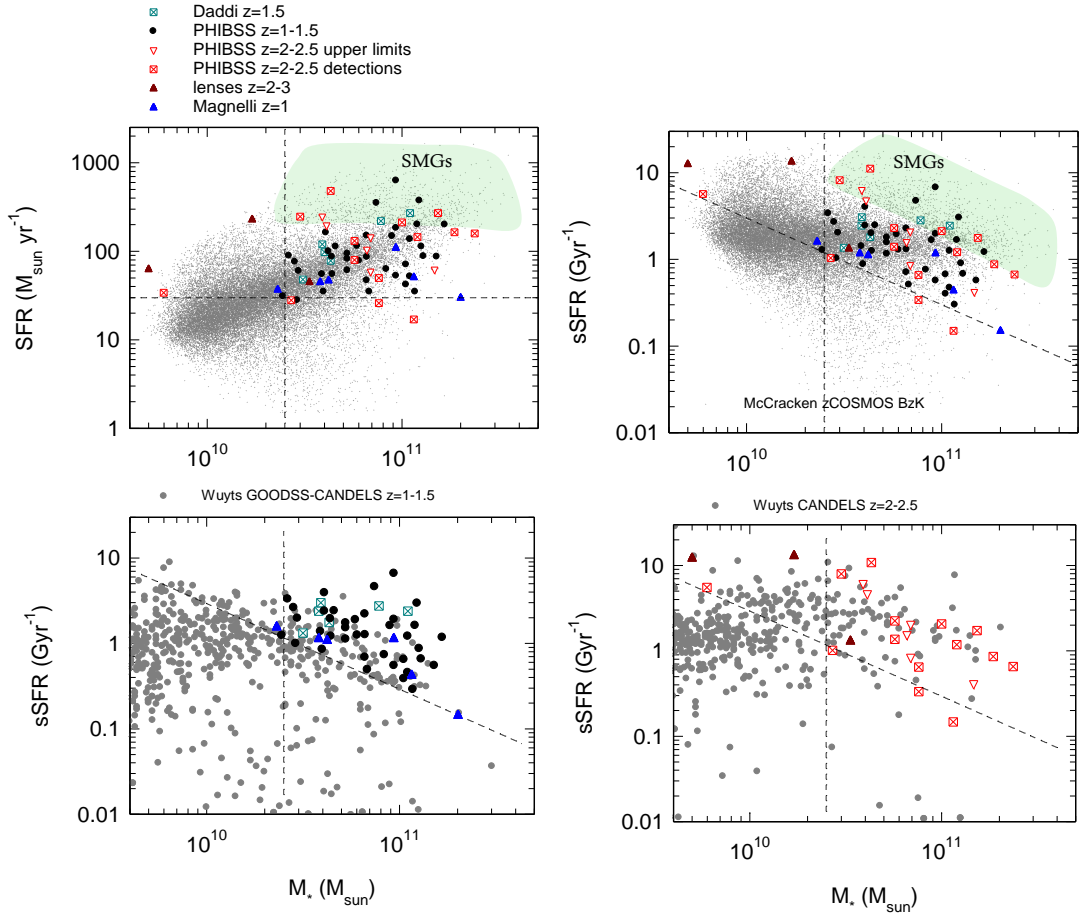

Figure 1. The location of the PHIBSS survey SFGs in the stellar mass – star formation and stellar mass- specific star formation planes. Filled black circles, open crossed red squares and open red triangles mark the basic PHIBSS survey detections at z=1-1.5, the detections at z=2-2.5 and the non-detections (<3 σ) at z=2-2.5. Filled blue triangles denote the six Herschel-PACS-selected detections at z=1-1.5 by Magnelli et al. (2012b). Open crossed green squares denote z~1.5 SFG detections from Daddi et al. (2010a), and brown filled triangles the three lenses cB58 (z=2.7, Baker et al. 2004), the 'cosmic eye": (z=3.1, Coppin et al. 2007) and the 'eyelash' (z=2.3, Swinbank et al. 2010, Danielson et al. 2011). The vertical and horizontal/diagonal dotted black lines denote the stellar mass and star formation cuts

in our survey ($M_* \geq 2.5 \times 10^{10}$ M$_\odot$, $SFR \geq 30$ M$_\odot$ yr$^{-1}$). The green shaded region marks the location of z=1-4 submillimeter detected star forming galaxies (SMGs: $S_{850\mu m}$>3-10mJy, Magnelli et al. 2012a, Bothwell et al. 2012). Upper left: Survey data as described above in the stellar mass- star formation plane, along with the distribution of z~1.5-2.5 BzK SFGs in the COSMOS field (McCracken et al. 2010, Mancini et al. 2011, grey dots). Upper right: Survey data and BzK-COSMOS data in the stellar mass – specific star formation plane. Bottom: Comparison of survey data at z=1-1.5 (left) and z=2-2.5 (right) with the GOODS-S CANDELS data (grey filled circles) of Wuyts (priv.comm.). The best fit 'main-sequence' is given by $SFR$ (M$_\odot$ yr$^{-1}$)=45 ($M_*/6.6 \times 10^{10}$ M$_\odot$)$^{0.65}$((1+z)/2.2)$^{2.8}$, $sSFR$ (Gyr$^{-1}$)=0.68 ($M_*/6.6 \times 10^{10}$ M$_\odot$)$^{-0.35}$ ((1+z)/2.2)$^{2.8}$. Bouché et al. 2010, Noeske et al. 2007, Daddi et al. 2007, Rodighiero et al. 2010, Salmi et al. 2012, Whitaker et al. 2012, Wuyts priv.comm.).

**log M$_*$<10.7**

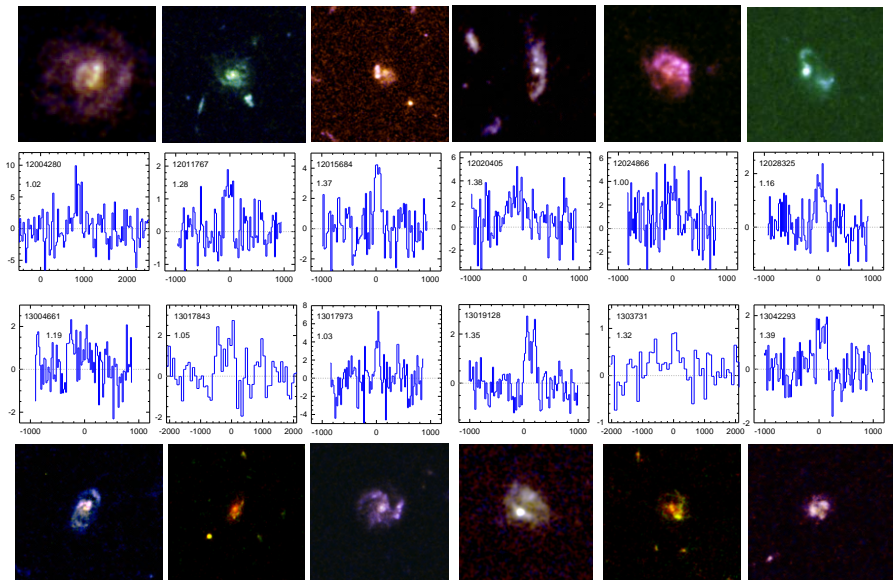

## 10.72 < log M* < 11.0

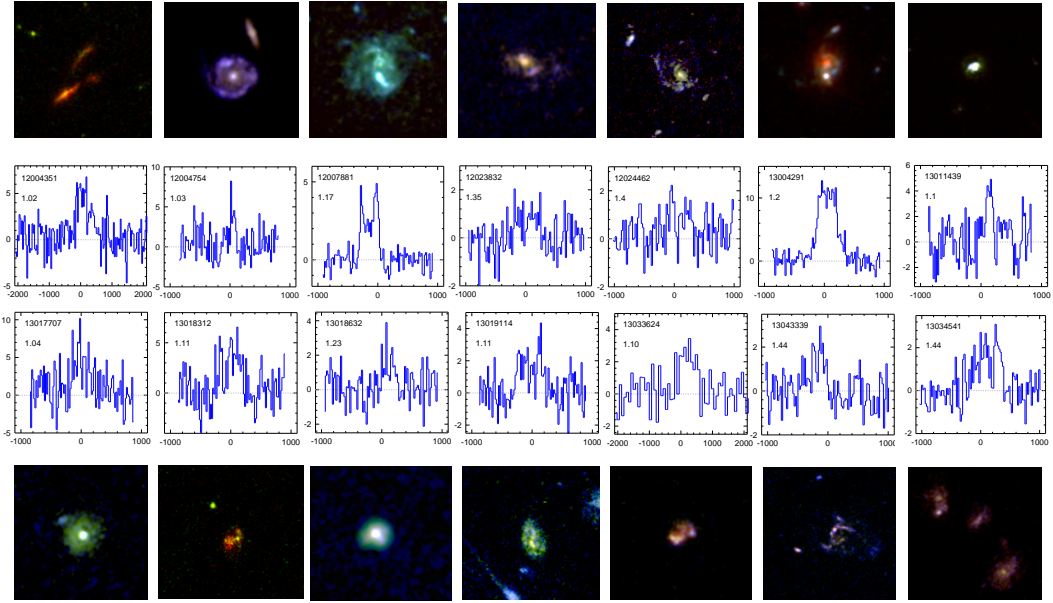

## log M*>11

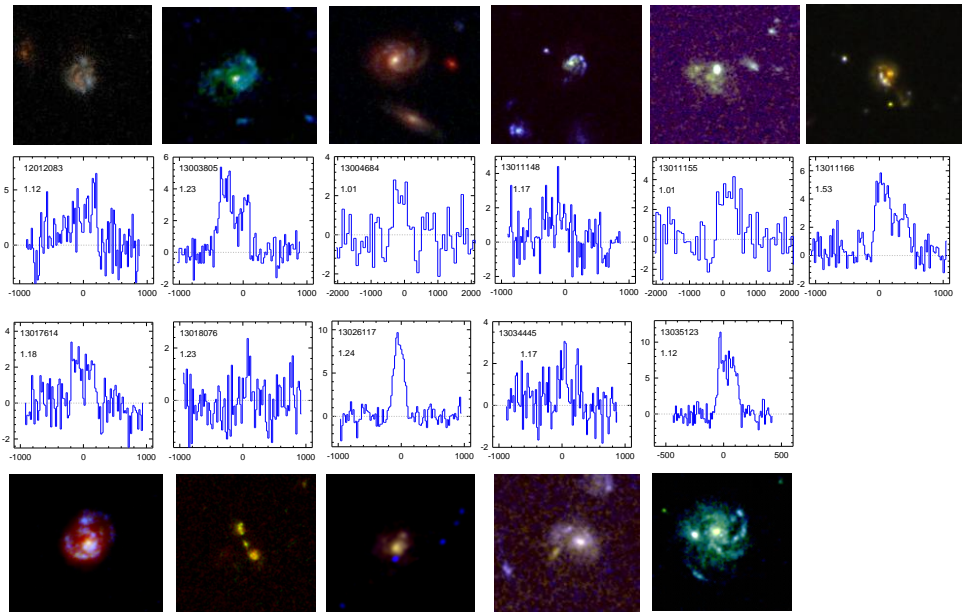

Figures 2-4: Galaxy integrated CO 3-2 spectra (vertical axis is flux density in mJy/beam, horizontal axis velocity offset (km/s) from systemic redshift given in the inset) and HST images (mostly in ACS I, V bands, some combining ACS I/V and CANDELS H) of the z=1-1.5 sample in three stellar mass bins: logM∗<10.7, 10.7≤logM∗<11.1, logM∗>11.1.

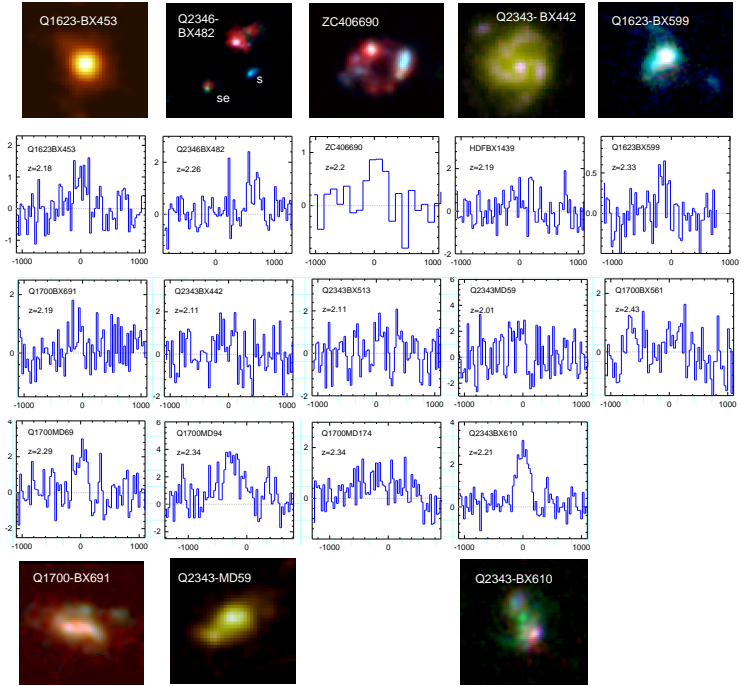

Figure 5. Galaxy integrated CO 3-2 spectra (vertical axis is flux density in mJy/beam, horizontal axis velocity offset (km/s) from systemic redshift given in the inset) for the z=2-2.5 detections (>3σ). The spectra are ordered in the same three bins of increasing stellar mass as in Figures 2-4, from top left to bottom right. Wherever available we also show WFC3-ACS, WFC3, or WFC3 + Hα composite images (from SINS/zC, Förster Schreiber et al. 2011, 2012, and from Law et al. 2012b). In the case of BX482 the redshift of the CO emission suggests that most of the emission comes from the fainter source BX482se, rather than from the main galaxy BX482 that dominates the rest-frame optical emission.

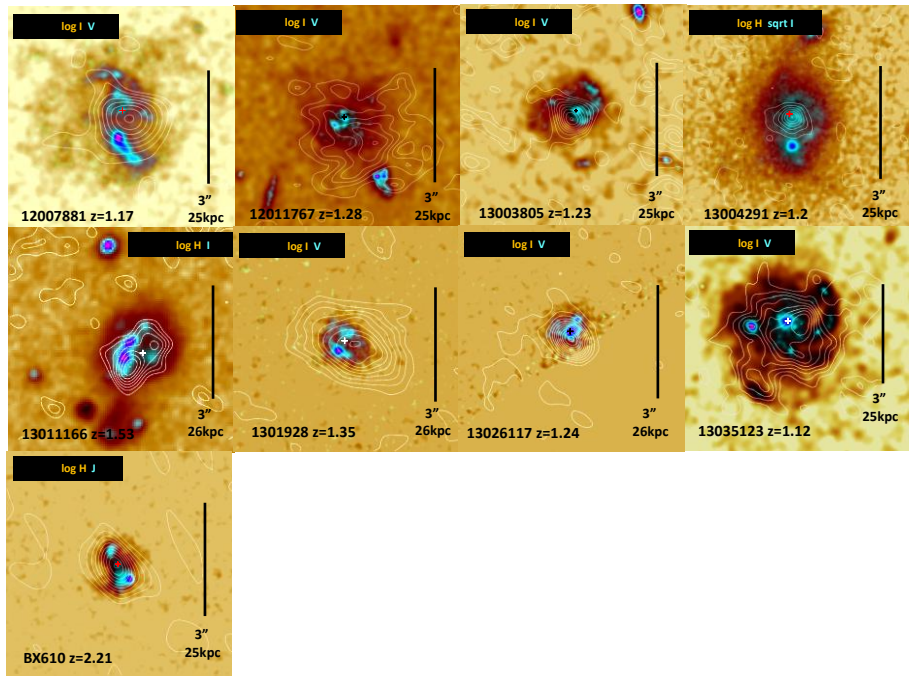

Figure 6: Superposition of CO high resolution maps (white contours, FWHM resolutions 0.4"x0.5" (EGS13004291) to 0.7"x1.5" (BX610)) on HST images (with bands labeled in each panel). All galaxy images are approximately on the same linear scale. The kinematics of these 9 galaxies ranges from ordered disk rotation (EGS12007881, 13003805, 13011166, 1301928, 13035123, Q2343 BX610), to compact SFG (EGS13026117), an interacting galaxy pair (12011767) and a late stage merger (EGS12004291).

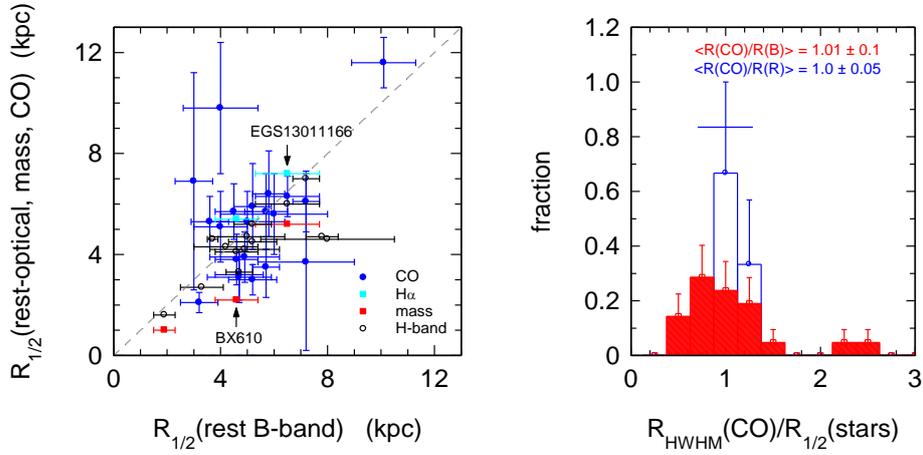

Figure 7. Comparison of source sizes in molecular gas and stellar tracers. Left: Comparison of derived half-light radii estimated from I-band data (~rest-B-band, horizontal axis) and H-band (~rest-R-band, open black circles), CO (filled blue circles), Hα (filled cyan squares) and stellar mass (filled red squares). The dotted grey line marks the location of $R_{1/2}(CO) = R_{1/2}(UV/optical)$. To the SFGs with spatially resolved CO emission in this paper, we also added 4 SFGs with CO sizes from Daddi et al. (2010a). Right: distribution of the ratio of CO to stellar sizes in rest-frame B-band (red) and rest-frame R-band (blue). The vertical error bars denote the Poisson uncertainties and the blue horizontal bar denotes the average ±1σ uncertainty of individual measurements of the ratio.

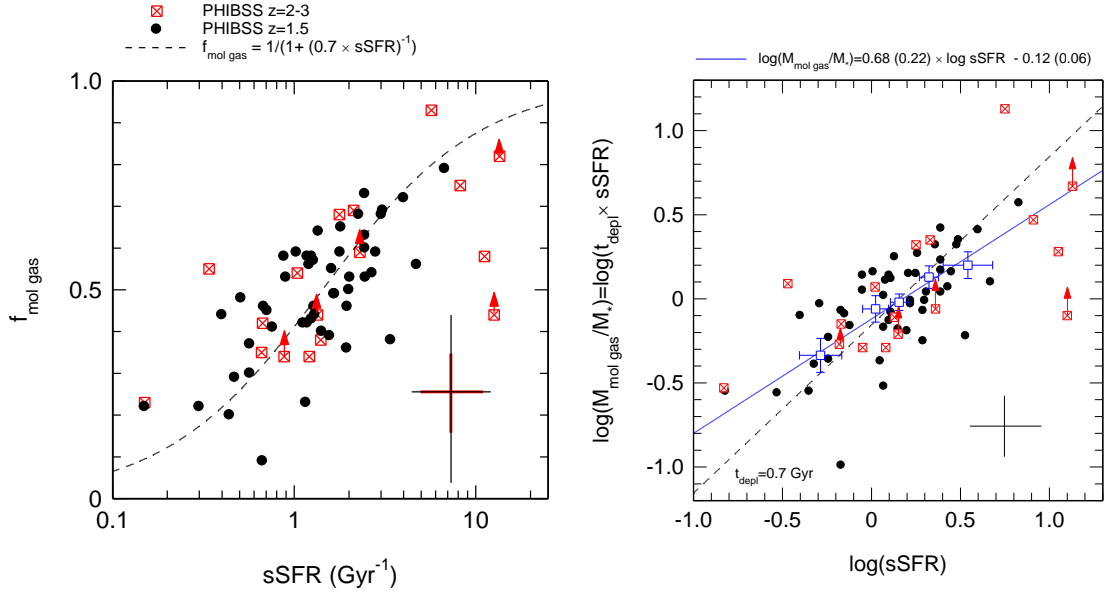

Figure 8. Left: Correlation between specific star formation rate (*sSFR=SFR/M*∗*) and molecular gas fraction, for the 50 PHIBSS z=1-1.5 SFGs (including Daddi et al. 2010a, +Magnelli et al. 2012b: filled black circles) and the z=2-2.5 detections (crossed red squares, arrows denoting possible corrections for metallicity). The dashed curve is the best fit relation of equation (3) with a constant depletion time scale of $7 \times 10^8$ years. Right: A different representation of the same relation, where we *plot log($M_{mol-gas}/M_*$) vs log(sSFR)*. Blue squares denote binned averages of the data, with horizontal and vertical bars denoting dispersion and uncertainty of the median. The solid line indicates the best fit with slope 0.7 (±0.2); points and dashed line are the same as in the left panel. The dispersion of these relations about the best fit with constant depletion time is ±0.24 dex.

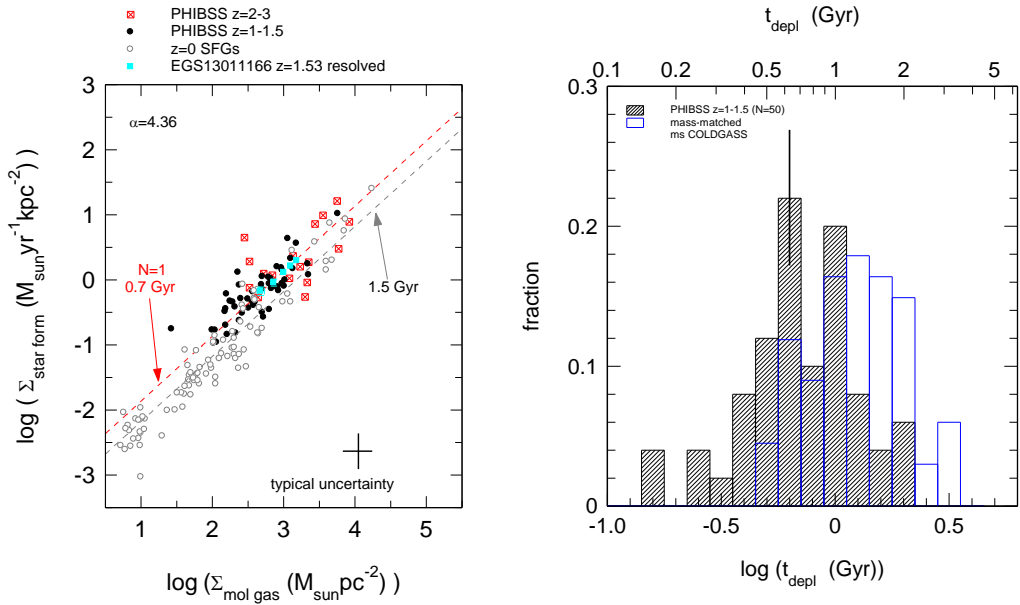

Figure 9: Molecular gas – star formation scaling relation. Left: Molecular gas – star formation surface density relationship ('Kennicutt-Schmidt' relation), for z~0 SFGs (open grey circles: Kennicutt 1998a, Gracia-Carpio 2008, Genzel et al. 2010, Armus et al. 2009), z=1-1.5 SFGs (black filled circles: PHIBSS, including Daddi et al. 2010a, Magnelli et al. 2012b), z=1.53 EGS13011166 (filled cyan squares: PHIBSS team, in prep.), z=2-2.5 detected SFGs (open crossed red squares, PHIBSS). The dotted grey and red lines mark the best fit linear (N=1) fits to the low-z and high-z data. Right: Distribution of depletion time scales inferred from the 50 z=1-1.5 SFGs in the left panel (black shaded histogram). The distribution has a median of $t_{depl}$=0.7 Gyr and a dispersion of 0.24 dex. For the SDSS selected, z~0 COLDGASS survey for main-sequence galaxies matched to the same mass range and coverage of the main-sequence as PHIBSS (Saintonge et al. 2011a), the corresponding depletion time scale

and scatter are 1.24±0.06 Gyr and ±0.23 dex (blue histogram). A typical Poisson error for both low- and high-z distributions is indicated.

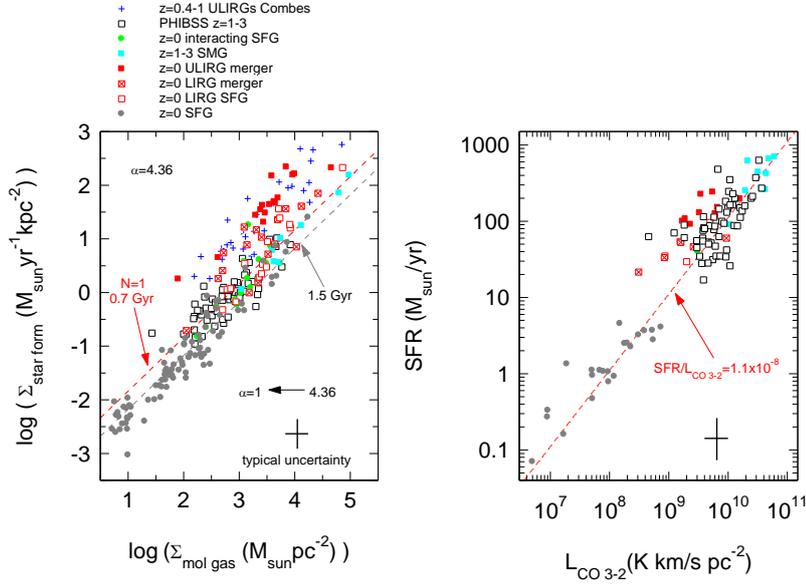

Figure 10. Left: Comparison of gas surface density-star formation surface density relation of the z=0 and z=1-3 main-sequence SFGs (grey filled circles and black open squares), with various off-main sequence populations, all for a common α=4.36 Milky Way, conversion factor. Open and crossed red squares denote z=0 single (non-merging: open red squares) and merging (crossed red squares) luminous infrared galaxies (LIRGs: $L_{IR} \sim 10^{11...12}$ $L_\odot$) from the GOALS survey (Armus et al. 2009), filled red squares denote ultra-luminous (ULIRGs: $L_{IR} \geq 10^{12}$ $L_\odot$) mergers in the Kennicutt (1998a), Gracia Carpio 2008 and GOALS (Armus et al. 2009) surveys. Blue crosses denote z=0.4-1 ULIRGs from Combes et al. (2012). Filled green circles denote more distantly interacting z~0 galaxies in the same surveys. Cyan filled squares mark z=1-4 submillimeter detected galaxies (SMGs) from the compilations of Tacconi et al. (2006, 2008), Engel (2010), Bothwell et al. (2012) and Magnelli et al.

(2012a). A change of the conversion factor from the Milky Way to the 'ULIRG' value ($\alpha\sim 1$) is marked by a horizontal black arrow. Right: Comparison of the observed CO luminosity in the J=3-2 line in the high-z SFGs and SMGs with z~0 SFGs, LIRGs and ULIRGs, with the nomenclature the same as in the left panel. The CO 3-2 observations in the local samples are from Iono et al. (2009), Wilson et al. (2012) and Papadopoulos et al. (2010).

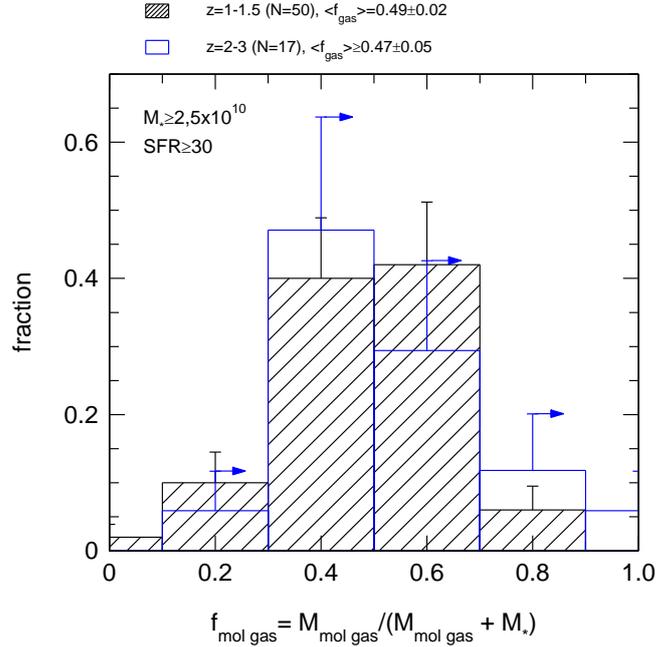

Figure 11. Molecular gas fractions for the z=1-1.5 (black hatched) and z=2-3 (blue) detected PHIBSS galaxies, for a Milky Way conversion factor ($\alpha$=4.36). For z=1-1.5 we combined the 38 SFGs from the sample presented in Tables 1 and 2, with the 6 z=1-1.5 detections of the 'PACS'-selected sample of Magnelli et al. (2012b) and the 6 z~1.5 detections of Daddi et al. (2010a). At z=2-3 we combined our 14 detections with the detections of three somewhat lower mass ($M_* = 5\text{-}30 \times 10^9 \, M_\odot$), strongly lensed SFGs between z~2.3 and 3.1 (cB58: z=2.7, Baker et al. 2004, 'cosmic eye": z=3.1, Coppin et al. 2007, 'eyelash': z=2.3, Swinbank et al. 2010, Danielson et al. 2010). 5 of the 14 z=2-2.5 SFGs have metallicities significantly lower than solar. In this regime the conversion factor probably is larger than the Milky Way value and the plotted gas fractions are lower limits (Genzel et al. 2012). The inferred median/average molecular gas fractions are given at the top of the figure.

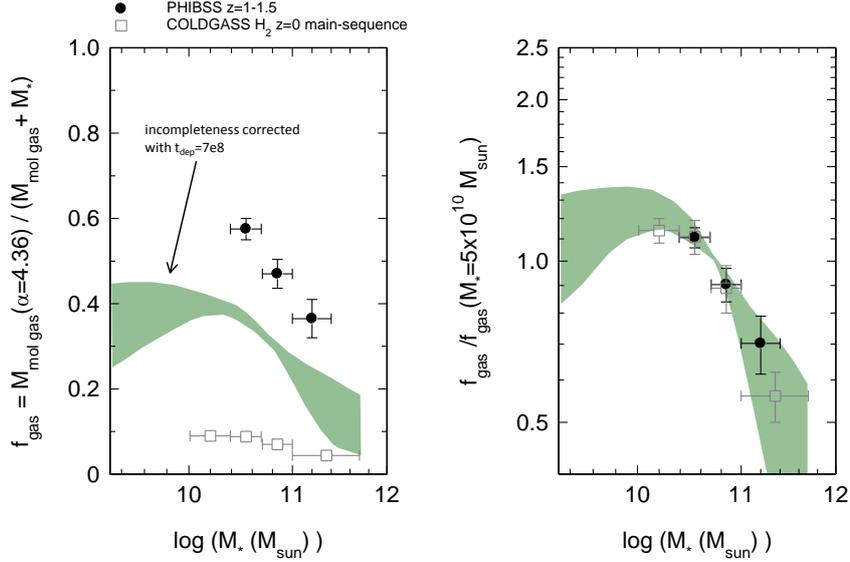

Figure 12: Molecular gas fractions at as a function of stellar mass. The left panel shows the observed median gas fractions, the right panel gas fractions normalized to the value at $M_*=5\times10^{10}$ $M_\odot$. The black circles are medians of the binned distribution of the 50 z~1-1.5 detected SFGs of our base sample, including the 12 z~1.5 detections from Daddi et al. (2010a) and Magnelli et al. (2012a), shown in Table 2. The grey open squares are averages of the z=0 SFGs from the COLDGASS survey for galaxies within 0.5dex of the main-sequence (Saintonge et al. 2011a). The bars in horizontal direction denote the bin-size, while the bars in vertical direction give the uncertainty in the mean of the points in each bin (~17 for each of the three bins). The green shaded area gives an estimate of the z=1-1.5 gas fractions corrected for the incompleteness of the PHIBSS survey in the area below the main sequence. For this purpose we took the dependence of specific star formation rate on stellar mass in various recent imaging surveys (*sSFR* (Gyr[-]

[1])=a*($M_*$/6.6x10^{10})^p ((1+z)/2.2)^q, with a=0.0.42 Gyr^{-1}, p=-0.1…-0.5, q~2.8, Noeske et al. 2007, Daddi et al. 2007, Panella et al. 2009, Rodighiero et al. 2010, Bouche et al. 2010, Mancini et al. 2011, Karim et al. 2011, Salmi et al. 2012, Whitaker et al. 2012, Wuyts priv. comm., see also Figure 1) and estimated gas fractions from equation (3), with $t_{dep}$=7x10$^8$ yr, as derived from the K-S relation in Figure 9.

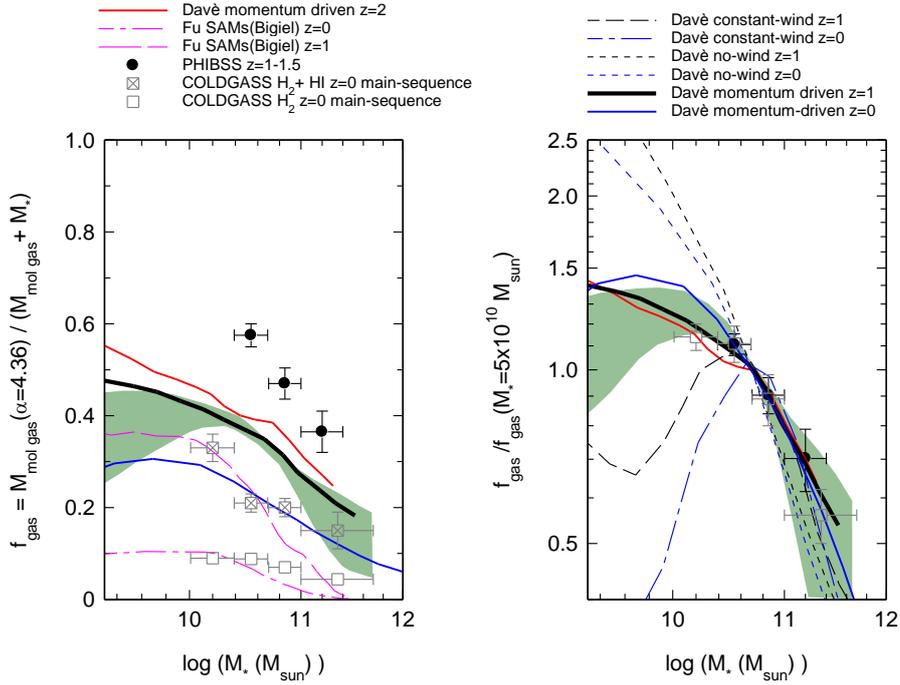

Figure 13: Comparison of the observed and incompleteness corrected z=1-1.5 molecular gas fractions, as a function of stellar mass, with the cosmological galaxy evolution models. The data and symbols are the same as in Figure 12, left and right panels again denote absolute and normalized values. The blue (for z=0), black (z=1) and red (z=2) solid lines denote the model predictions from cosmological hydrodynamical simulations with momentum driven winds by Davé et al. (2011, similar values were obtained by Guo et al. 2010, Ocvirk private communication). Dotted blue and black curves denote no wind models (at z=0 and 1), and blue and black dash-dotted curves constant wind velocity models (at z=0 and 1) from Davé et al. (2011). Dash-dotted and solid magenta curves denote semi-analytic models (at z=0 and 1) from Fu et al. (2012) that include supernova energy driven winds. The mass dependence of the momentum driven wind models matches the observations very well. This is particularly impressive for the normalized plots in the right panel, but also applies for the absolute values in the left panel if the observed z=1 gas fractions are corrected for incompleteness (green-shading), as discussed in Figure 12, and if the combined HI + molecular gas fractions are taken at z=0 (crossed grey squares, from Saintonge et al. 2011a, b). The models do not distinguish between molecular and atomic gas.

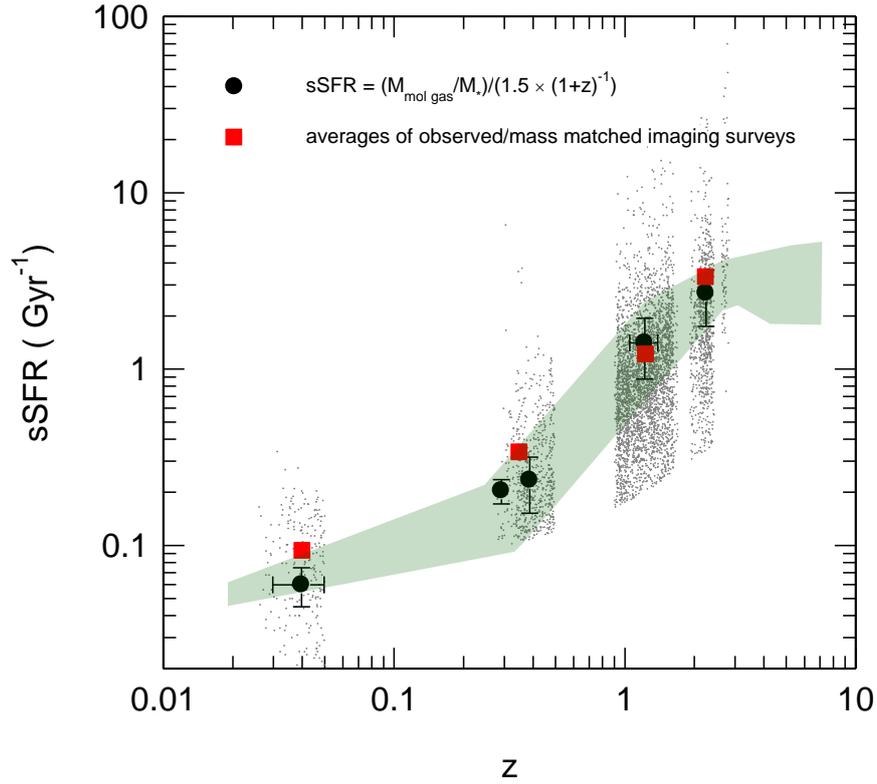

Figure 14. Specific star formation rate *sSFR* ($Gyr^{-1}$) as a function of redshift. The large black filled circles denote the specific star formation rates inferred from the average CO gas fractions, with *sSFR = ($M_{mol\ gas}/M_*$)/$t_{depl}$*. We adopt $t_{depl}=M_{mol\ gas}/SFR=1.5 \times (1+z)^{-1}$ Gyr inferred from the COLDGASS and PHIBSS surveys (1.5 Gyr at z=0 and 0.7 Gyr at z=1.2). The COLDGASS average and dispersion ($<M_{gas}/M_*>$=0.085±0.02) are for the same mass selection as for PHIBSS ($<M_{gas}/M_*>$=0.89±0.34 at $<z>$=1.23 and $<M_{gas}/M_*>$=1.17±0.41 at $<z>$=2.27). The z=0.39 point is an average of 5 above main-sequence SFGs, with stellar masses ranging between 5 and $10 \times 10^{10}$ $M_\odot$ (Geach et al.2011). The z=0.294 point is an average of 5 near-main sequence SFGs of $logM_*$~11.3 from Bauermeister et al. (2012). The grey dots are individual galaxies from SDSS and imaging surveys (Saintonge et al. 2011a, Wuyts et al. 2011 and priv.comm.) mass-matched to the CO-surveys, and the filled red squares denote the average *sSFR* derived from these data points in the four redshift slices of the CO surveys. The grey-green shaded region are estimates of mean *sSFR* as a function of redshift from various imaging surveys in the literature, as compiled in Weinmann et al. (2011, with references therein), Sargent et al. (2010, with references therein) and Gonzalez et al (2012). The points from the Sargent compilation lie at the lower end of the shaded distribution and refer to stellar masses comparable to the CO surveys ($M_*$~$5 \times 10^{10}$ $M_\odot$), including for the SDSS point from Brinchmann et al. (2004) at z~0.02.

**Table 1: CO Observations**

| Source | Configs | $t_{int}$ (hours) | RA optical | Dec Optical | RA CO peak | Dec CO peak | $z_{CO}$ | CO beam size |
|---|---|---|---|---|---|---|---|---|
| EGS12004280 | D | 5.2 | 14h17m00.9s | 52d27'01.3" | 14h17m00.9s | 52d27'01.3" | 1.023 | 2.8"x2.5" |
| EGS12004351 | D | 5.2 | 14h17m02.0s | 52d26'58.5" | 14h17m01.8s | 52d27'00.2" | 1.017 | 2.8"x2.4" |
| EGS12004754 | C | 5.4 | 14h16m42.1s | 52d25'19.1" | 14h16m42.0s | 52d25'18.4" | 1.026 | 1.6"x1.4" |
| EGS12007881 | ABCD | 9.4 | 14h18m03.6s | 52d30'22.2" | 14h18m03.5s | 52d30'22.6" | 1.160 | 1.1"x0.9" |
| EGS12011767 | ACD | 7.7 | 14h18m24.8s | 52d32'55.4" | 14h18m24.7s | 52d32'54.8" | 1.282 | 2.0"x1.6" |

NOTE: This table is published in its entirety in the electronic edition of the Astrophysical Journal. The first rows are shown here to illustrate the form and content.

## Table 2: Derived Properties

| Source | Type[1] | $v_{rot}$[2] km/s | $R_{1/2opt}$[3] Kpc | $R_{1/2CO}$[3] kpc | SFR[4] $M_\odot$/yr | F(CO)[5] Jy km/s | σ(CO) Jy km/s | L(CO)[5] K km/s pc$^2$ | $M_{mol-gas}$[6] $M_\odot$ | $M_*$[7] $M_\odot$ | $f_{gas}$[8] | log$\Sigma_{molgas}$[9] $M_\odot$/pc$^2$ | log$\Sigma_{SFR}$[9] $M_\odot$/yr/kpc$^2$ |
|---|---|---|---|---|---|---|---|---|---|---|---|---|---|
| EGS12004280 | Disk(A) | 154 | 4.7 | --- | 100 | 1.24 | 0.24 | 7.9E+09 | 7.0E+10 | 4.1E+10 | 0.63 | 2.70 | -0.14 |
| EGS12004351 | Merger | 302 | 5.7 | 5.7 | 150 | 2.53 | 0.34 | 1.6E+10 | 1.4E+11 | 8.0E+10 | 0.64 | 2.84 | -0.14 |
| EGS12004754 | Disk(A) | 69 | 6.5 | --- | 53 | 0.73 | 0.07 | 4.7E+09 | 4.1E+10 | 9.3E+10 | 0.31 | 2.19 | -0.70 |
| EGS12007881 | Disk(A) | 220 | 5.7 | 5.6 | 94 | 1.03 | 0.07 | 8.5E+09 | 7.5E+10 | 5.2E+10 | 0.59 | 2.57 | -0.34 |
| EGS12011767 | Int. disk | 147 | 5.8 | 6.4 | 76 | 0.38 | 0.03 | 3.8E+09 | 3.3E+10 | 2.8E+10 | 0.54 | 2.20 | -0.44 |

[1]: see section 3.1

[2]: rotation velocity, typical uncertainty 20-30%

[3]: half light radius, taken from HST I-band fitting for AEGIS and Hα for BX (column 5), and from CO 3-2 (column 6), typical uncertainty ±25%

[4]: extinction corrected star formation rate from 24μm+ UV continuum for AEGIS and Hα+UV for BX, Chabrier IMF, systematic uncertainty ±35%

[5]: CO 3-2 (or 2-1 for BzK and PEP) integrated line flux and integrated line luminosity; minus-sign stands for a 3σ upper limit

[6]: molecular gas mass, corrected for helium (1.36), a 'Galactic' CO-H2 conversion factor (X=2x10$^{20}$ cm$^{-2}$/ (K km/s)), or α=4.36, and a CO1-0/CO3-2 ratio of 2, systematic uncertainty 50%

[7]: stellar mass from SED fitting, assuming a Chabrier IMF, systematic uncertainty ±30%

[8]: $f_{gas} = M_{mol-gas}/(M_{mol-gas} + M_*)$

[9]: $\Sigma_{mol\,gas} = 0.5\,M_{mol\,gas}/\pi(R_{1/2})^2$, $\Sigma_{star\,form} = 0.5\,SFR/\pi(R_{1/2})^2$

NOTE: This table is published in its entirety in the electronic edition of the Astrophysical Journal. The first rows are shown here to illustrate the form and content.

# Table 3: Ratio of $v_{rot}/\sigma_0$ in high-z SFGs

|  | $<v_{rot}>$ km/s | $<\sigma_0>$ km/s | $<v_{rot}/\sigma_0>$ | References |
|---|---|---|---|---|
| Hα IFU z=1-2.5, logM∗>10.4 (N=18) | 237 (±37) | 50 (±16) | 4.7 (±1.8) | Genzel et al. 2008, Förster Schreiber et al. 2009, 2012, Epinat et al. 2012 |
| cosmic eyelash z=2.33 CO 1-0/6-5 | 320 (±25) | 50 (±10) | 6.4 (±1.4) | Swinbank et al. 2011 |
| EGS12007881 z=1.17 CO 3-2 | 232 (±46) | 32 (±6) | 7.3 (±2.0) | this paper, Combes et al. (2012b) |
| EGS13003805 z=1.23 CO 3-2 | 350 (±71) | 46 (±11) | 7.8 (±2.2) | this paper, Combes et al. (2012b) |
| EGS13004291 z=1.197 CO 3-2 | 260 (±66) | 51 (±8) | 5.1 (±1.5) | this paper, Combes et al. (2012a) |
| EGS13011166 z=1.53 CO 3-2 | 370 (±60) | 55 (±8) | 6.7 (±1.5) | this paper, MPE et al. 2012 |
| EGS13011166 z=1.53 Hα | 370 (±60) | 65 (±15) | 5.7 (±1.6) | this paper, MPE et al. 2012 |
| EGS13019128 z=1.35 CO 3-2 | 170 (±40) | 28 (±8) | 6.1 (±2) | this paper, Combes et al. (2012b) |
| EGS13035123 z=1.12 CO 3-2 | 210 (±30) | 22 (±6) | 9.5 (±3) | this paper, Tacconi et al. (2010) |